\documentclass[twocolumn]{aastex61}

\usepackage{amsmath}
\usepackage{mhchem}

\received{}
\revised{}
\accepted{}
\submitjournal{ApJ}

\shorttitle{Chemical properties of two dense cores in PGCC G168.72-15.48}
\shortauthors{Tang et al.}

\begin{document}
\title{Chemical properties of two dense cores in a Planck Galactic Cold Clump G168.72-15.48}

\AuthorCollaborationLimit=200

\author[0000-0001-9160-2944]{Mengyao Tang}
\affil{Department of Astronomy, Yunnan University, and Key Laboratory of Astroparticle Physics of Yunnan Province, Kunming, 650091, China, mengyao\_tang@yeah.net, slqin@bao.ac.cn}

\author{J. X. Ge}
\affiliation{Departamento de Astronom\'{\i}a, Universidad de Chile, Camino el Observatorio 1515, Las Condes, Santiago, Chile}

\author{Sheng-Li Qin}
\affiliation{Department of Astronomy, Yunnan University, and Key Laboratory of Astroparticle Physics of Yunnan Province, Kunming, 650091, China, mengyao\_tang@yeah.net, slqin@bao.ac.cn}

\author{Tie Liu}
\affiliation{Shanghai Astronomical Observatory, Chinese Academy of Sciences, 80 Nandan Road, Shanghai 200030, People’s Republic of China}

\author{Yuefang Wu}
\affiliation{Department of Astronomy, Peking University, 100871, Beijing China}

\author{Kee-Tae Kim}
\affiliation{Korea Astronomy and Space Science Institute, 776 Daedeokdaero, Yuseong-gu, Daejeon 34055, Republic of Korea}
\affiliation{University of Science and Technology, 217 Gajeong-ro, Yuseong-gu, Daejeon 34113, Republic of Korea }

\author{Sheng-Yuan Liu}
\affiliation{Institute of Astronomy and Astrophysics, Academia Sinica. 11F of Astronomy-Mathematics Building, AS/NTU No. 1, Section 4, Roosevelt Rd., Taipei 10617, Taiwan}

\author{Chao Zhang}
\affiliation{Department of Astronomy, Yunnan University, and Key Laboratory of Astroparticle Physics of Yunnan Province, Kunming, 650091, China, mengyao\_tang@yeah.net, slqin@bao.ac.cn}
\affiliation{Department of Astronomy, Peking University, 100871, Beijing China}

\author{J. H. He}
\affiliation{Key Laboratory for the Structure and Evolution of Celestial Objects, Yunnan Observatories, Chinese Academy of Sciences, P.O. Box 110, Kunming, 650011, Yunnan Province, China}
\affiliation{Chinese Academy of Sciences, South America Center for Astrophysics (CASSACA), Camino El Observatorio 1515, Las Condes, Santiago, Chile}
\affiliation{Departamento de Astronom\'{i}a, Universidad de Chile, Las Condes, Santiago, Chile}

\author{Bing-Gang Ju}
\affiliation{Purple Mountain Observatory, Qinghai Station, 817000, Delingha, China}

\author{Xinhe Fang}
\affiliation{Department of Electronic Engineering, Zhengzhou Railway Vocational \& Technical College, Zhengzhou, 451460, China}


\begin{abstract}
To deepen our understanding of the chemical properties of the Planck Galactic Cold Clump (PGCC) G168.72-15.48, we performed observations of nine molecular species, namely, \ce{c-C3H}, \ce{H2CO}, \ce{HC5N}, \ce{HC7N}, \ce{SO}, \ce{CCH}, \ce{N2H+}, \ce{CH3OH}, and \ce{CH3CCH}, toward two dense cores in PGCC G168.72-15.48 using the Tianma Radio Telescope and Purple Mountain Observatory Telescope.
We detected \ce{c-C3H}, \ce{H2CO}, \ce{HC5N}, \ce{N2H+}, \ce{CCH}, and \ce{CH3OH} in both G168-H1 and G168-H2 cores, whereas \ce{HC7N} and \ce{CH3CCH} were detected only in G168-H1 and SO was detected only in G168-H2.
Mapping observations reveal that the \ce{CCH}, \ce{N2H+}, \ce{CH3OH}, and \ce{CH3CCH} emissions are well coupled with the dust emission in G168-H1.
Additionally, \ce{N2H+} exhibits an exceptionally weak emission in the denser and more evolved G168-H2 core, which  may be attributed to the \ce{N2H+} depletion. We suggest that the \ce{N2H+} depletion in G168-H2 is dominated by \ce{N2} depletion, rather than the destruction by CO.
The local thermodynamic equilibrium calculations indicate that the carbon-chain molecules of \ce{CCH}, \ce{HC5N}, \ce{HC7N}, and \ce{CH3CCH} are more abundant in the younger G168-H1 core.
We found that starless core G168-H1 may have the properties of cold dark clouds based on its abundances of carbon-chain molecules. While, the prestellar core G168-H2 exhibits lower carbon-chain molecular abundances than the general cold dark clouds.
With our gas-grain astrochemical model calculations, we attribute the observed chemical differences between G168-H1 and G168-H2 to their different gas densities and different evolutionary stages.
\end{abstract}


\section{Introduction}
\label{Introduction}
\begin{figure}
\plotone{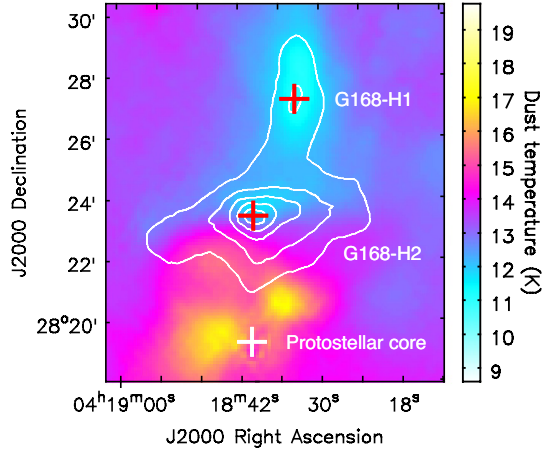}
\caption{\emph{Herschel} \ce{H2} column density map (white contours) overlaid on the \emph{Herschel} dust temperature map (color). White contours range from 30\% to 90\%, with a step of 15\% of peak value ($4.48\times10^{22}$ cm$^{-2}$). Red crosses represent the peak positions of the G168-H1 and G168-H2 cores. White cross denotes a point source IRAS 04155+2812 \citep{Ward-Thompson16}.}
\label{Herschel&JCMT}
\end{figure}

Planck Galactic Cold Clumps (PGCCs) have been considered as good targets for investigating the early stage of star formation.
Many observations have been conducted to understand the properties of PGCCs \citep{Liu12,Meng13,Zhang16,Wu12,Liu14,Liu18}.
For example, L1495, a filamentary cloud harboring active low-mass star forming regions, contains 16 PGCCs \citep{Planck16}.
Recent observations of 16 PGCCs by \cite{Tang18} identified 30 dense cores in L1495, and classified these dense cores into three groups: starless, prestellar, and protostellar cores.
G168.72-15.48 (hereafter denoted as G168) is one of the 16 PGCCs detected in L1495. Figure~\ref{Herschel&JCMT} presents the spatial distributions of the dust temperature and \ce{H2} column density derived from the \emph{Herschel} continuum imaging data of G168. As seen from Figure~\ref{Herschel&JCMT}, there are two dense cores, G168-H1 and G168-H2. A protostellar core IRAS 04155+2812 (marked with a white cross in Figure~\ref{Herschel&JCMT}) located south of G168-H2 is believed to be heating G168-H2 \citep{Ward-Thompson16}.
\citet{Dapp09}  proposed an \ce{H2} density profile to model spherical and flattened clouds, which is characterized by a flat region ($a_{d}$), central column density ($N_{\rm c}$), central volume density ($n_{\rm c}$), and truncated radius ($R_{\rm d}$).
The fitting \ce{H2} density profiles to the 30 dense cores are summarized in Table~\ref{Tab_para}.
As shown in Table~\ref{Tab_para}, the density profile parameters ($a_{\rm d}$, $N_{\rm c}$, $n_{\rm c}$) of G168-H1 and G168-H2 correspond to the statistical results of starless core and prestellar core, respectively. Therefore, \citet{Tang18} classified G168-H1 as a starless core and G168-H2 as a prestellar core. Starless cores will become prestellar cores when they are dense enough to be gravitationally bound. Then, the prestellar cores can collapse to form protostars \citep{Ward-Thompson94, Caselli11}. Under this sketch, G168-H2 is more evolved than G168-H1 during the earliest stage of star formation.

Considering their different evolutionary stages, G168-H1 and G168-H2 are good targets for studying properties of starless and prestellar cores.
Previous observations toward G168 by \citet{Tang18} mainly focused on the physical properties, demonstrating that G168-H1 and G168-H2 have source sizes of $394^{\prime\prime}\times158^{\prime\prime}$ (P.A.=$-6^{\circ}$) and $251^{\prime\prime}\times169^{\prime\prime}$ (P.A.=$-85^{\circ}$) on the angular scale, corresponding to geometric mean core radii of 0.17($\pm$0.03) and 0.14($\pm$0.02) pc at a distance of 140 pc, respectively. Volume densities of 1.89($\pm$0.21)$\times10^{4}$ and 4.78($\pm$0.71)$\times10^{4}$ cm$^{-3}$, dust temperatures of 11.90($\pm$0.31) and 13.25($\pm$1.02) K, and core masses of 26.47($\pm$3.04) and 37.79($\pm$5.51) M$_{\sun}$ were also reported for G168-H1 and G168-H2, respectively (see Table~\ref{Tab_para}).
However, the chemical properties of these dense cores in G168 have not been well investigated.
In this study, we investigated these two cores to deepen our understanding of the chemical properties of starless and prestellar cores in G168.

In cold dark clouds, carbon-chain and nitrogen-bearing molecules can be easily detected  and are generally used for probing the properties of cold dark clouds \citep{Benson83,Benson98,Suzuki92,Scappini96,Tafalla98,Aikawa01,Hirota04,SakaiN08,SakaiN10,Suzuki14,Taniguchi18}.
PGCCs are mainly embedded in dense regions within cold dark clouds \citep{Planck16}.
Therefore, nitrogen-bearing and carbon-chain molecules are effective for revealing the chemical properties of PGCCs.
In this paper, we present the observations of nine molecular species, namely, \ce{HC5N}, \ce{HC7N}, \ce{c-C3H}, \ce{H2CO}, \ce{SO}, \ce{CCH}, \ce{N2H+}, \ce{CH3CCH}, and  \ce{CH3OH}, to investigate the chemical properties of  two dense cores in G168, through the observations by use of the Purple Mountain Observatory (PMO) 13.7 m telescope and TianMa 65 m Radio Telescope (TMRT).

\begin{deluxetable*}{cccccccccccccc}
\tabletypesize{\tiny}
\tablewidth{0pt}
\tablecaption{Parameters of G168-H1 and G168-H2}
\tablehead{
\colhead{Source} &Angle size                         &Radius &$N_{\rm H2}$             &$T_{\rm d}$ &$n_{\rm H2}$         &Mass   &$X_{\rm C^{18}O}$      &$a_{\rm d}$$^{(a)}$ &$N_{\rm c}$$^{(a)}$           &$n_{\rm c}$$^{(a)}$           \\
\colhead{}       &($\prime\prime\times\prime\prime$) &(pc)   &(10$^{22}$ cm$^{-2}$)    &(K)         &(10$^{4}$ cm$^{-3}$) &(M$_{\sun}$)  &(10$^{-7}$)  &(pc)        &(10$^{22}$ cm$^{-2}$) &(10$^{4}$ cm$^{-3}$)
}
\startdata
G168-H1    &394$\times$158  &0.17($\pm$0.03) &1.40($\pm$0.21)  &11.90($\pm$0.31) &1.89($\pm$0.21) &26.47($\pm$3.04) &1.26($\pm$0.21) &0.047($\pm$0.017) &2.10($\pm$0.01) &4.65($\pm$1.75)   \\
G168-H2    &251$\times$169  &0.14($\pm$0.02) &2.46($\pm$0.62)  &13.25($\pm$1.02) &4.78($\pm$0.71) &37.79($\pm$5.51) &0.91($\pm$0.29) &0.025($\pm$0.001) &4.50($\pm$0.01) &17.74($\pm$1.3)   \\
\cline{1-12}
Starless   & & & & & & & & &  & &        \\
Mean       &...             &0.15($\pm$0.06) &0.86($\pm$0.27)  &12.11($\pm$0.41) &1.47($\pm$0.53) &15.90($\pm$15.21) &1.57($\pm$0.65) &0.048($\pm$0.006) &1.27($\pm$0.11) &3.50($\pm$0.55)   \\
Median     &...             &0.12($\pm$0.04) &0.86($\pm$0.26)  &11.13($\pm$0.41) &1.57($\pm$0.47) &7.73($\pm$1.40) &1.61($\pm$0.34) &0.047($\pm$0.004) &1.2($\pm$0.08)  &3.11($\pm$0.35)   \\
\cline{1-12}
Prestellar & & & & & & & & & & &  \\
Mean       &...             &0.10($\pm$0.03) &1.47($\pm$0.46)  &11.62($\pm$0.81) &3.96($\pm$1.05) &12.9($\pm$11.40) &0.81($\pm$0.13) &0.026($\pm$0.001) &2.49($\pm$0.02) &10.33($\pm$0.68)   \\
Median     &...             &0.10($\pm$0.02) &1.29($\pm$0.25)  &12.13($\pm$0.33) &4.43($\pm$0.94) &8.63($\pm$1.30) &0.84($\pm$0.11) &0.025($\pm$0.001) &2.10($\pm$0.01) &9.76($\pm$0.58)    \\
\enddata
\label{Tab_para}
\tablecomments{(a): $a_{\rm d}$, $N_{\rm c}$, and  $n_{\rm c}$ are flat region, central column density, and central volume density respectively. They were derived from the column density profile fitting toward the dense cores of L1495 dark cloud by \citet{Tang18}.  }
\end{deluxetable*}

\section{Observations}

\subsection{TMRT Observations}
We performed single-pointing observations toward the G168-H1 and G168-H2 cores in the Ku band (11.5--18.5 GHz) in August 2018 with the Tianma 65 m Radio Telescope (TMRT) of Shanghai Astronomical Observatory.
The observations were conducted in the position-switching mode.
The digital backend system (DIBAS) of TMRT supports several observing modes \citep{Bussa12}.
We adopted mode 20 for Ku band observations, which has a 187.5 MHz bandwidth with 32,768 channels,
providing a spectral resolution of 5.722 KHz (corresponding to a velocity resolution of 0.12 km s$^{-1}$) at the Ku band.
DIBAS supports two banks (A and B) for observing different frequency ranges simultaneously.
In our observations, the frequency ranges of bank A and B were set at 13--14 GHz and 14--15 GHz, respectively.
The transitions of \ce{HC5N}(5-4), \ce{HC7N}(11-10), \ce{HC7N}(12-11), and SO(1$_{2}$-1$_{1}$) were tuned in bank A, and the transitions of \ce{c-C3H}(\emph{J}=$\frac{1}{2}$-$\frac{3}{2}$), \ce{c-C3H}(\emph{J}=$\frac{3}{2}$-$\frac{3}{2}$), \ce{H2CO}(2$_{1,1}$-2$_{1,2}$), and \ce{HC7N}(13-12) were tuned in bank B.
An integrated time of 100 minutes on-source yielded a root-mean-square (rms) noise level of 20 mK for bank A. However, bank B was not stable enough during our observations, resulting in a higher rms noise level of 80 mK.
The beam size of TMRT ranges from 73$^{\prime\prime}$ to 54$^{\prime\prime}$ at the frequency range of 13--18 GHz. The average beam size is 69$^{\prime\prime}$, corresponding to a linear scale of $\sim$0.05 pc at a distance of 140 pc.
The main beam efficiency is 60\% at the Ku band \citep{Li16}.
The observations pointed at the peak positions of G168-H1 and G168-H2 at R.A.(J2000) = 4$^{\rm h}$18$^{\rm m}$33$^{\rm s}$.568, Decl.(J2000) = +28$^{\circ}$26$^{\prime}$56$^{\prime\prime}$805 and at R.A.(J2000) = 4$^{\rm h}$18$^{\rm m}$38$^{\rm s}$.298, Decl.(J2000) = +28$^{\circ}$23$^{\prime}$22$^{\prime\prime}$.959, respectively (corresponding to the red crosses in Figure~\ref{Herschel&JCMT}). The pointing accuracy was better than 12$^{\prime\prime}$. The observed molecular line data were processed using the GILDAS software package \citep{Guilloteau00}.

\subsection{PMO Observations}
Mapping observations of the \ce{CCH}, \ce{N2H+}, \ce{CH3OH}, and \ce{CH3CCH} lines toward G168 were performed with the PMO 13.7 m telescope in April 2018.
The nine-beam array receiver system, which operates at 85--115 GHz in double-sideband mode, was used as the frontend.
The fast Fourier transform spectrometer (FFTS) backend has 16,384 channels with a bandwidth of 1.0 GHz for each sideband.
The beam size of the telescope is approximately 53$^{\prime\prime}$, corresponding to a linear scale of $\sim$0.04 pc at a distance of 140 pc.  The main beam efficiency is 50\%.
We conducted on-the-fly mapping observations centered at R.A.(J2000)=04$^{\rm h}$18$^{\rm m}$34.58$^{\rm s}$, Decl.(J2000)=+28$^{\circ}$26$^{\prime}$34.64$^{\prime\prime}$.
A mapping region of 18$^{\prime}$$\times$18$^{\prime}$ in size was continuously scanned with a scan speed of 20$\arcsec$~s$^{-1}$.
The pointing accuracy was better than 5$^{\prime\prime}$.
The OTF data were processed by the GILDAS software package \citep{Guilloteau00}, and all data were converted to 3D cubes with a grid-spacing of 30$^{\prime\prime}$. The resulting spectral rms noise was approximately 80 mK.
The MIRIAD and CASA software packages were employed for further imaging and analysis \citep{Sault95,McMullin07}.
In this work, we only focused on the central 10$^{\prime}$ $\times$10$^{\prime}$ area, as the map edges have higher rms noise.

\section{Results}
\label{Results}
\subsection{TMRT results}
\begin{figure*}
\gridline{\fig{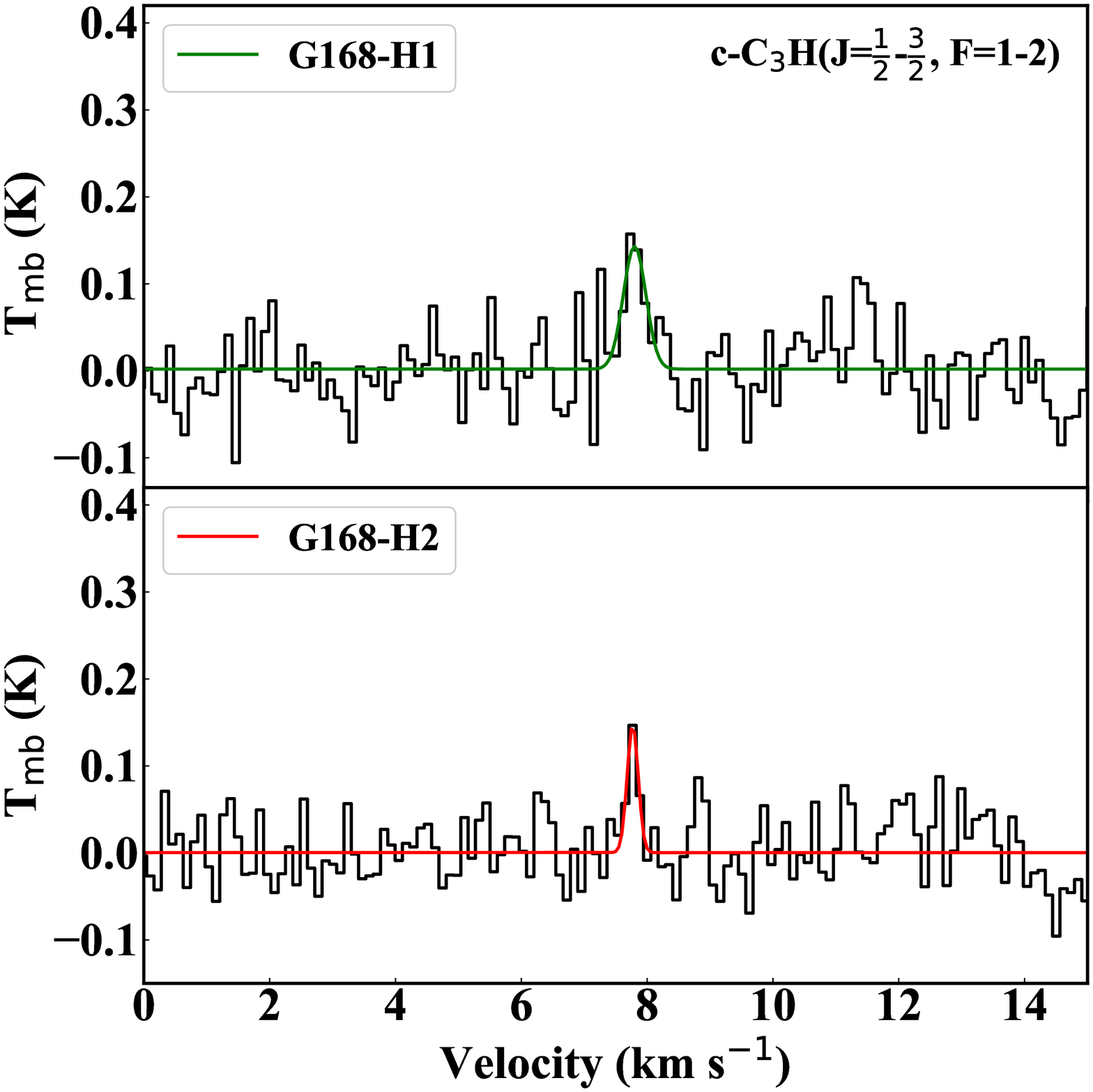}{0.33\textwidth}{(a)}
          \fig{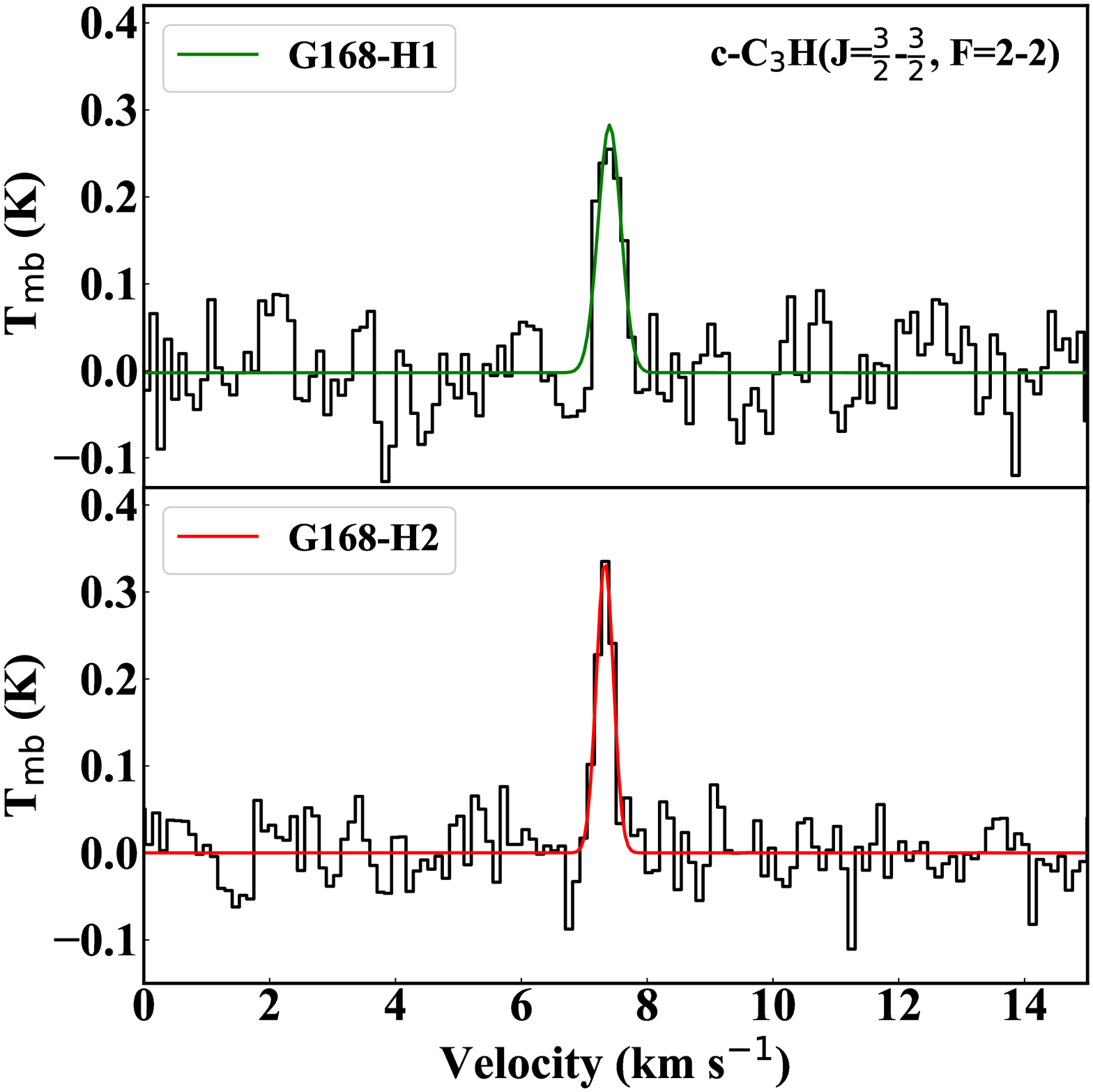}{0.33\textwidth}{(b)}
          \fig{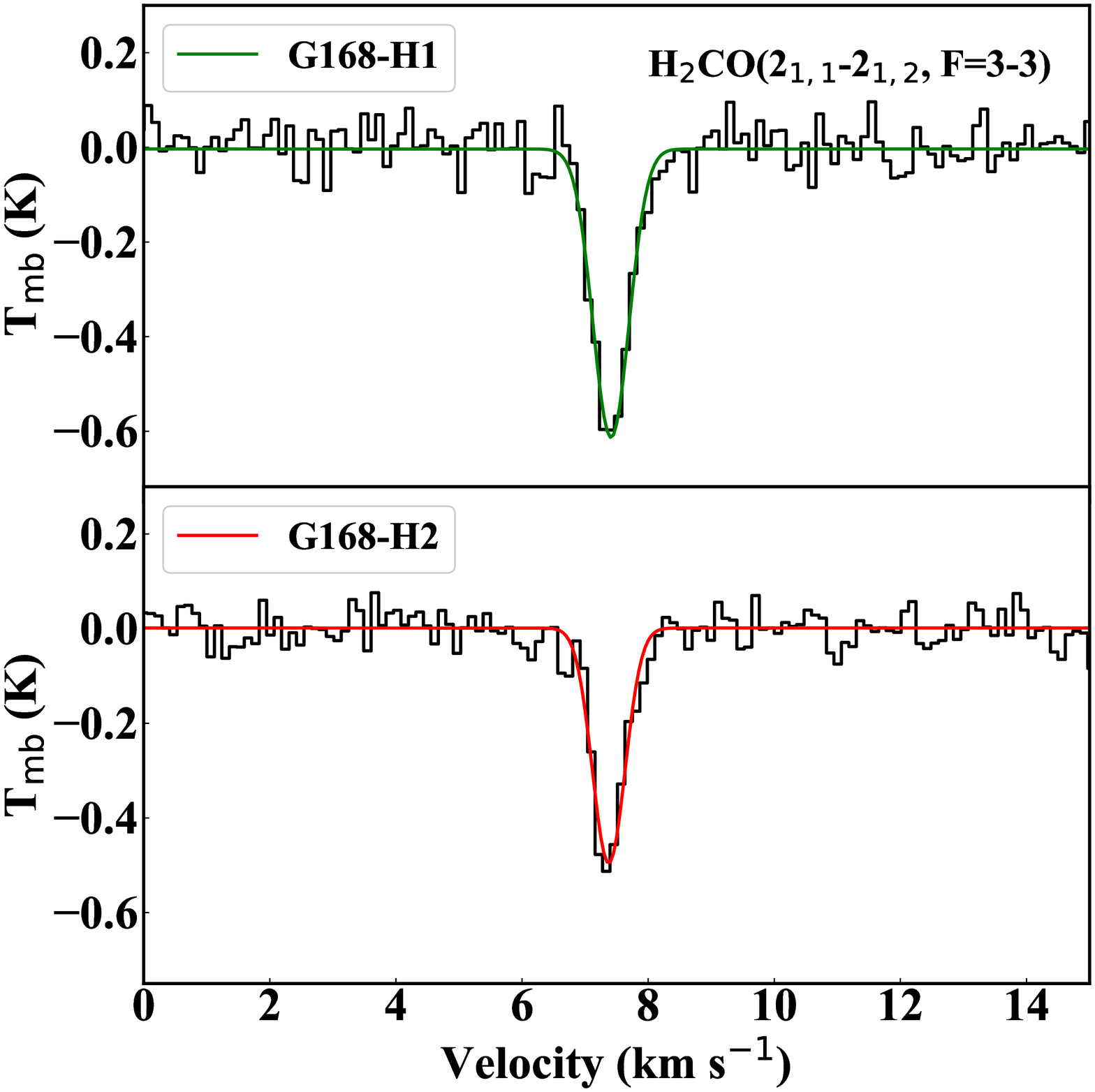}{0.33\textwidth}{(c)}
          }
\gridline{\fig{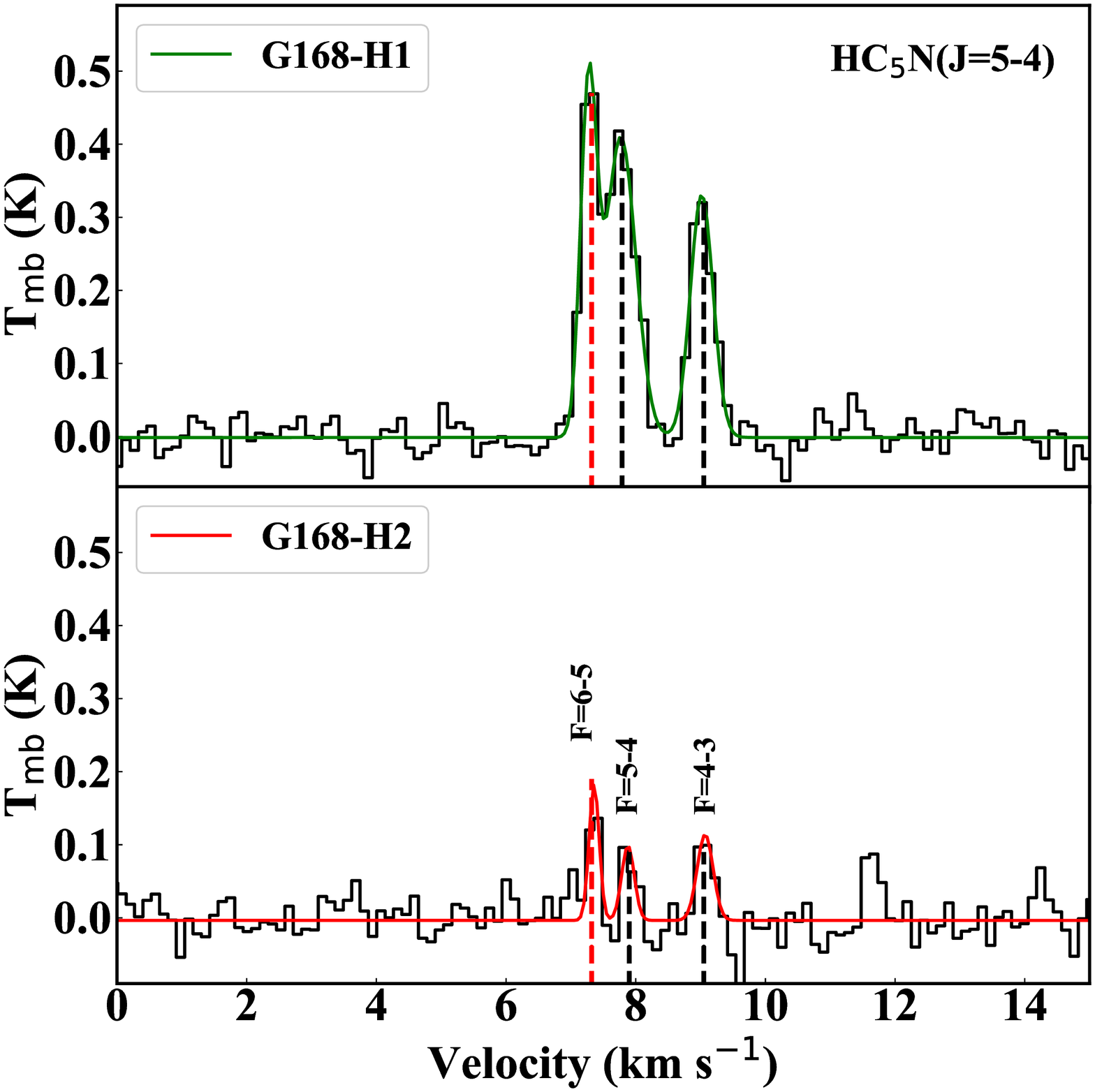}{0.33\textwidth}{(d)}
          \fig{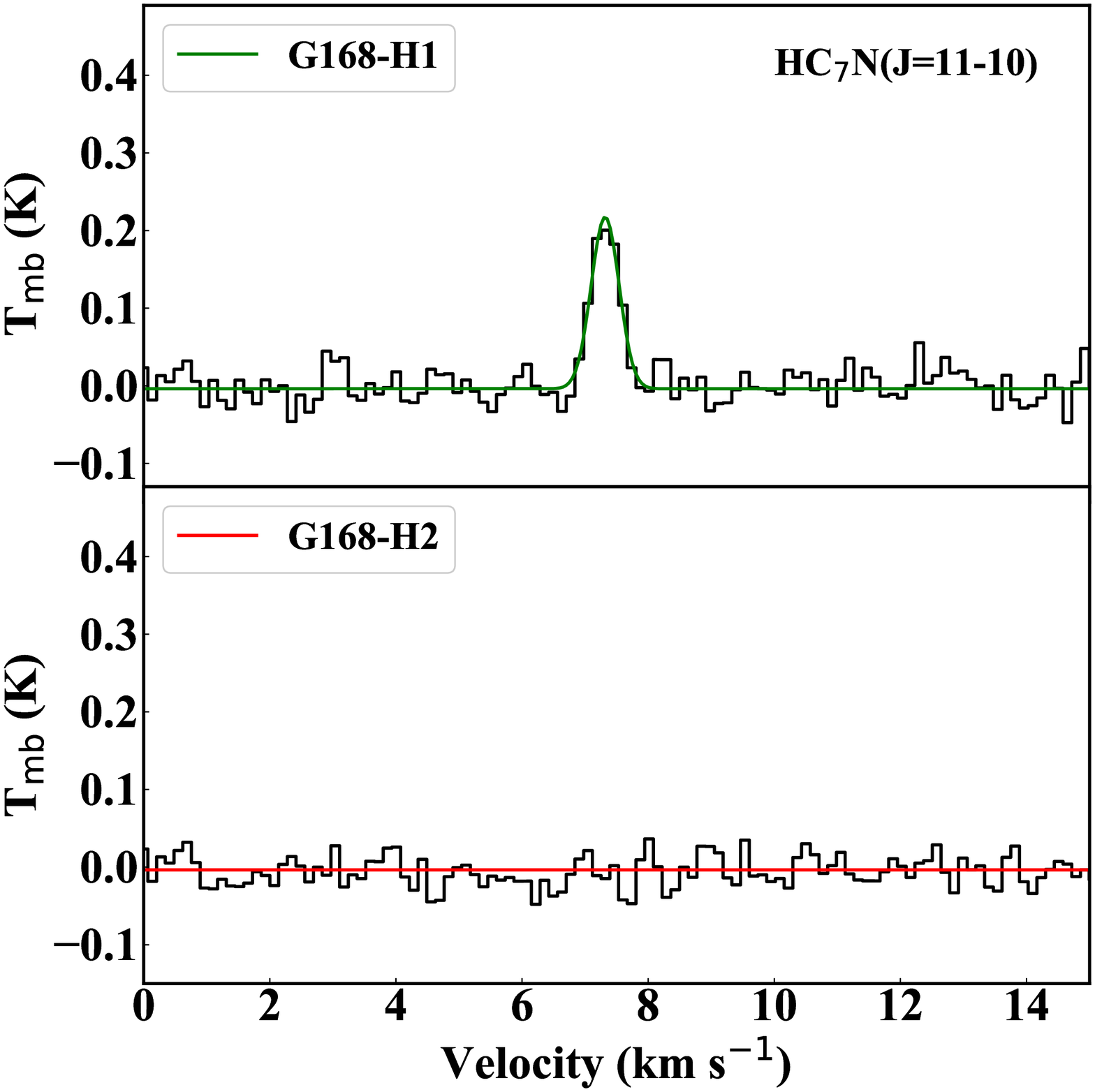}{0.33\textwidth}{(e)}
          \fig{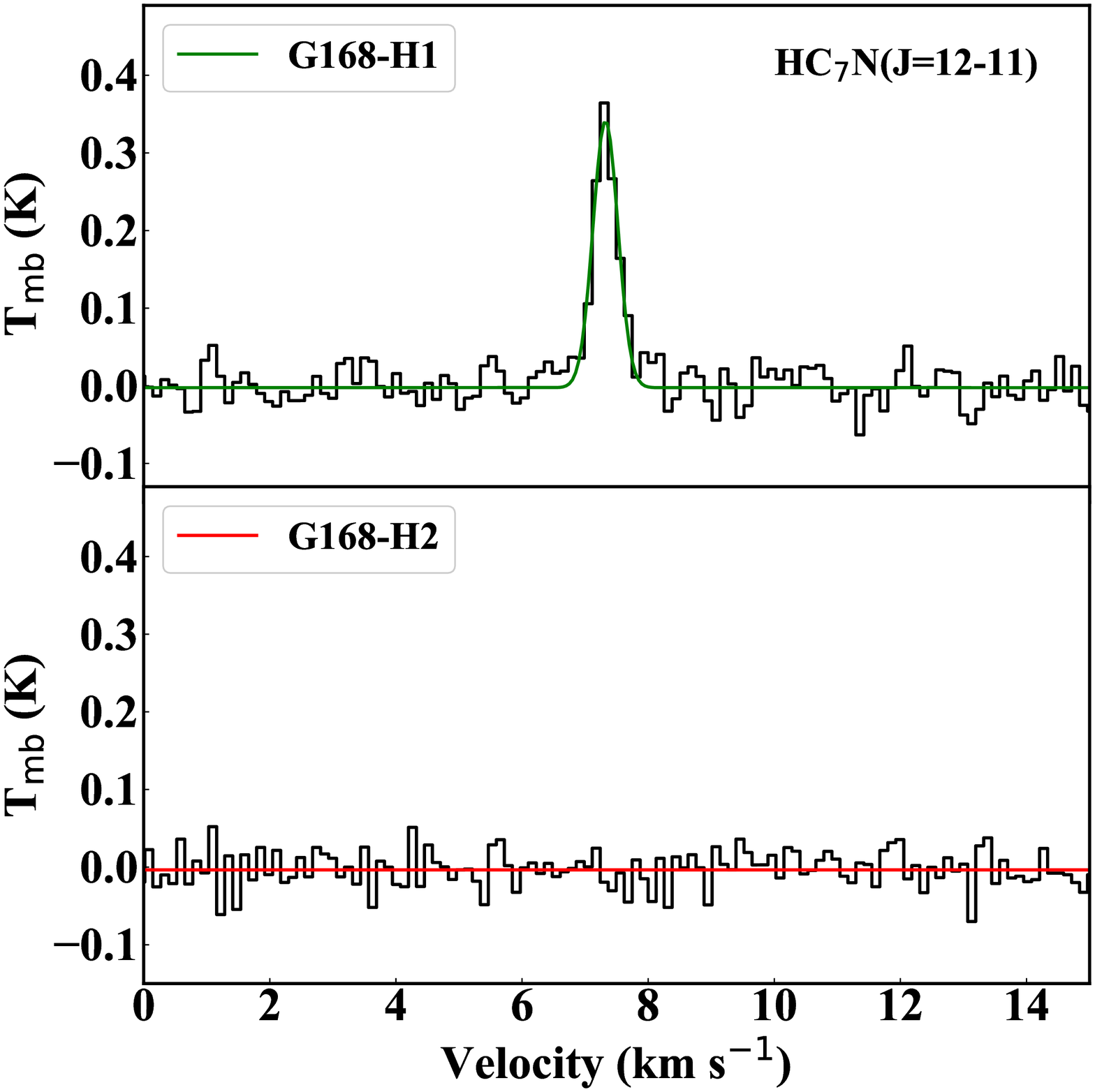}{0.33\textwidth}{(f)}
          }
\gridline{\fig{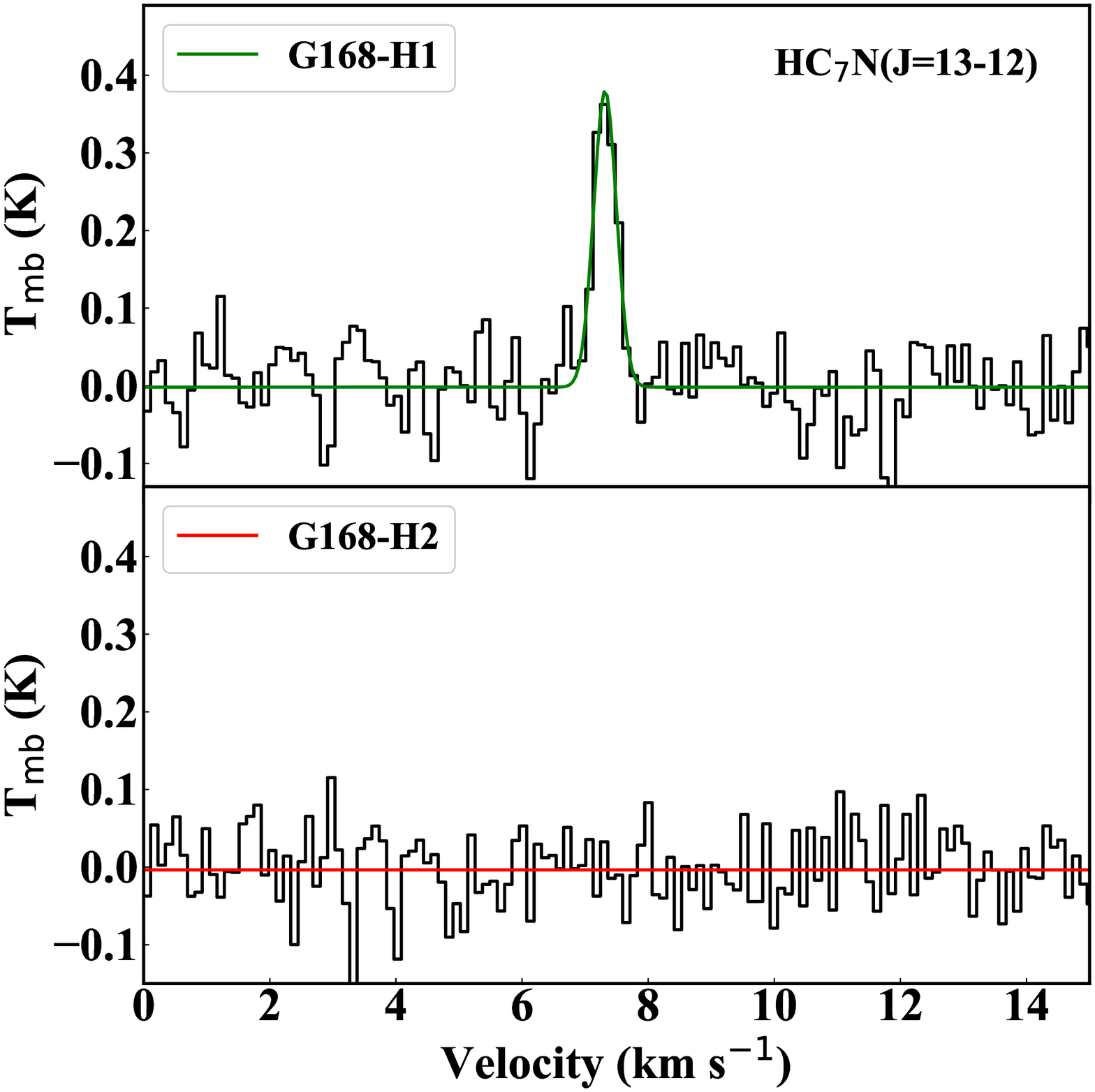}{0.33\textwidth}{(g)}
          \fig{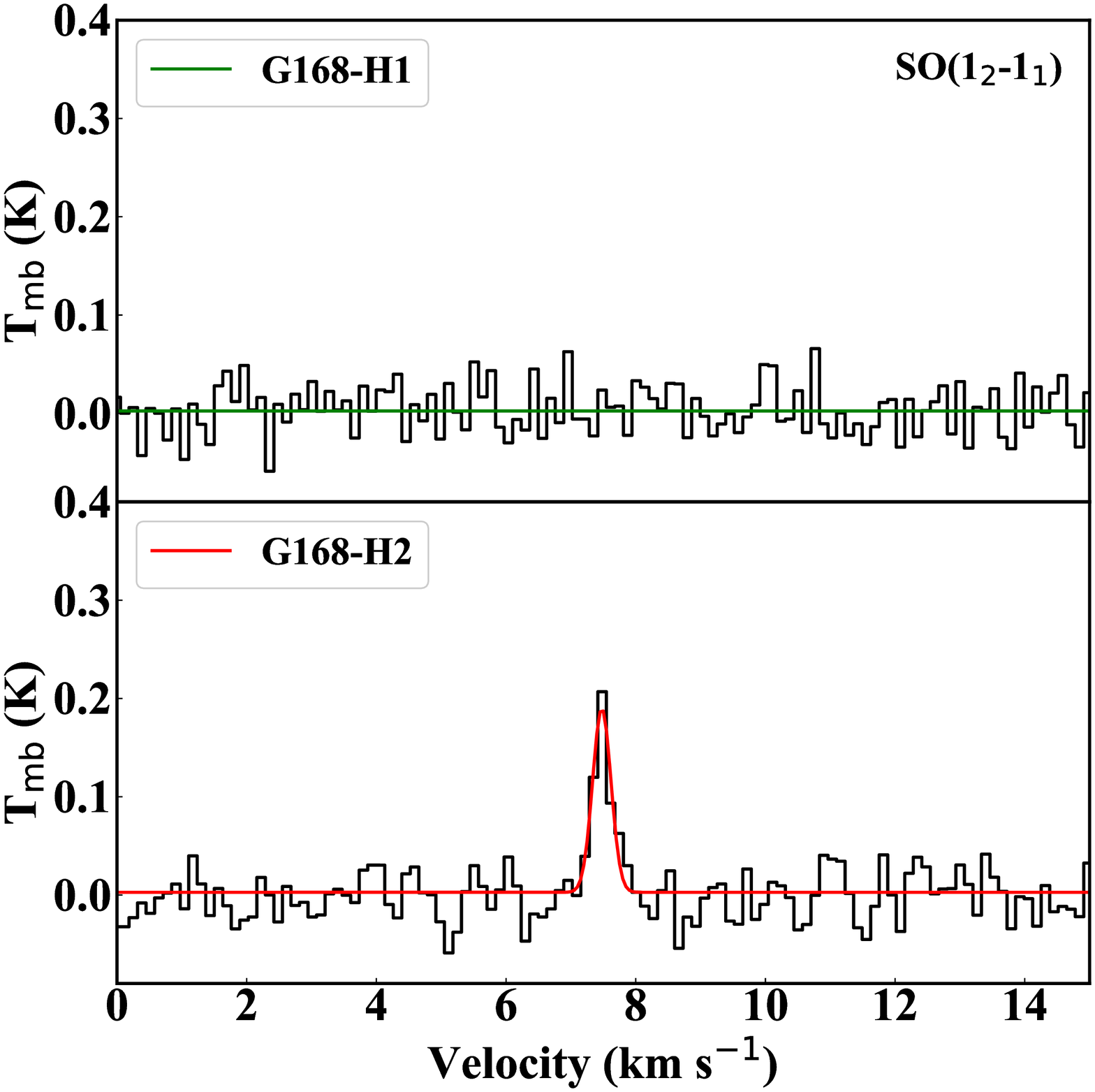}{0.33\textwidth}{(h)}
          }
\caption{Black step lines in the upper and bottom panels represent spectra observed by TMRT at the G168-H1 and G168-H2 peak positions, respectively. The green and red solid lines are the Gaussian fitting results of G168-H1 and G168-H2, respectively. The hyperfine structures are indicated by vertical dashed lines, and the red vertical dashed line denotes the velocity component corresponding to the systemic velocity.}
\label{TMRT_spec}
\end{figure*}

In total, two transitions of \ce{c-C3H}, one \ce{H2CO} absorption line, three hyperfine components of \ce{HC5N}, three transitions of \ce{HC7N}, and one SO line were observed by the TMRT observations.
The detailed information of these transitions is listed in Table~\ref{Tab_base}.
Figure~\ref{TMRT_spec} depicts the molecular spectra observed by TMRT.
The black step lines in the upper and bottom panels represent the spectra detected in the G168-H1 core and G168-H2 core, respectively.
As can be seen from panels (a) - (d) of Figure~\ref{TMRT_spec}, \ce{c-C3H}, \ce{H2CO}, and \ce{HC5N} were detected in both G168-H1 and G168-H2, and the line intensities of \ce{c-C3H} and \ce{H2CO} in the G168-H1 and G168-H2 cores are comparable. From panel (d) of Figure~\ref{TMRT_spec}, three hyperfine structure (HFS) lines of \ce{HC5N}(5-4) were resolved with its mean peak intensities of 0.47($\pm$0.05) K and 0.20($\pm$0.15) K in the G168-H1 and G168-H2 cores, respectively.
Whereas three transitions of \ce{HC7N} were detected only in G168-H1, the SO line was detected only in G168-H2.
Gaussian fitting was performed for extracting the line parameters, and the fitting results are presented in Table~\ref{Tab_base}. Note that the linewidth of each species is different, and then we chose a different velocity range for each species to perform Gaussian fitting. The Gaussian fitting results of G168-H1 and G168-H2 are indicated by the green and red lines in Figure~\ref{TMRT_spec}, respectively.

\subsection{PMO results}
In total, our PMO observations have detected three \ce{CCH} transitions, seven \ce{N2H+} transitions, two \ce{CH3OH} transitions, and two \ce{CH3CCH} transitions.
The detailed information of the observed transitions is listed in Table~\ref{Tab_base}.

\begin{figure*}
	\gridline{\fig{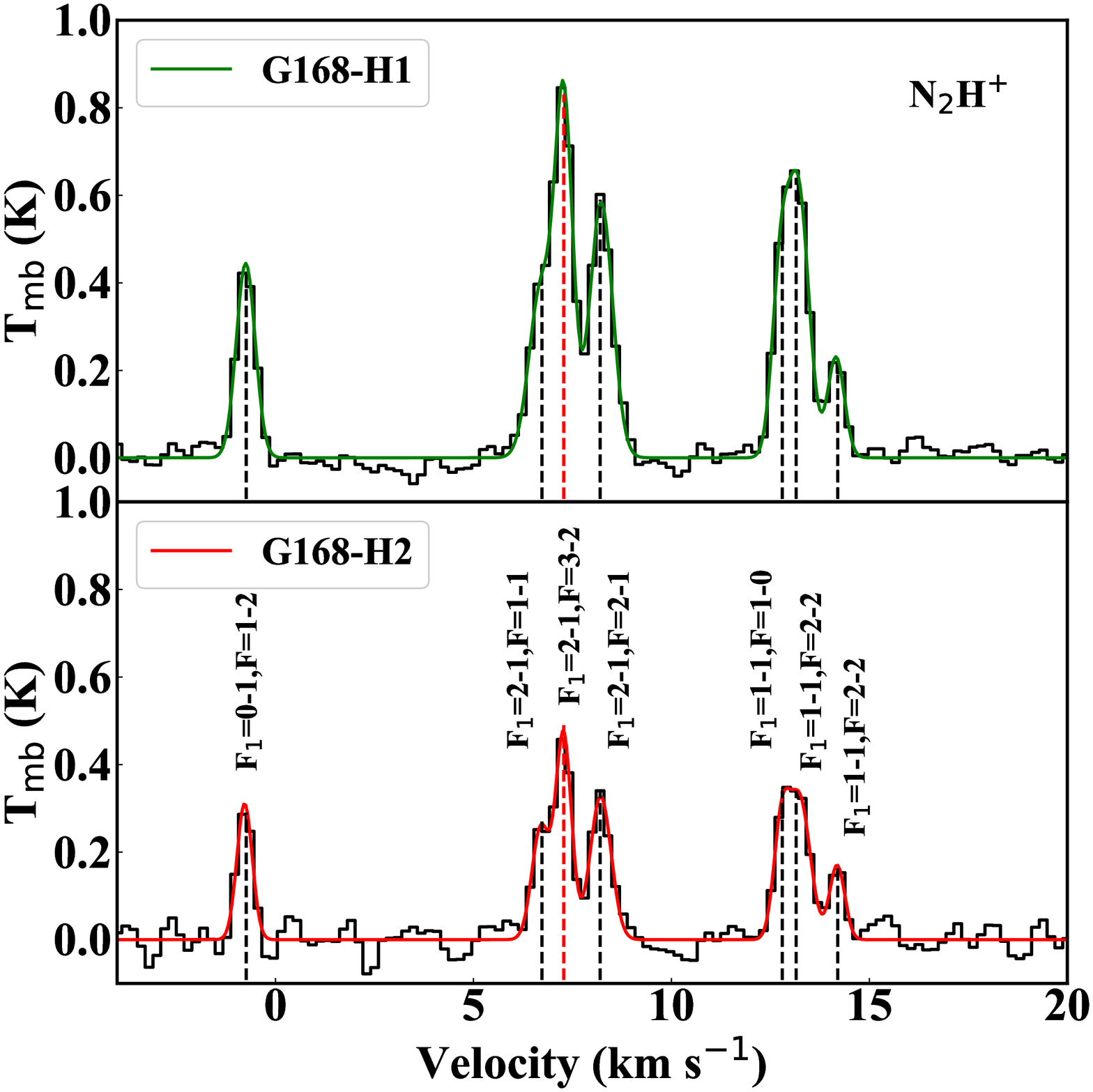}{0.3\textwidth}{(a)}
		\fig{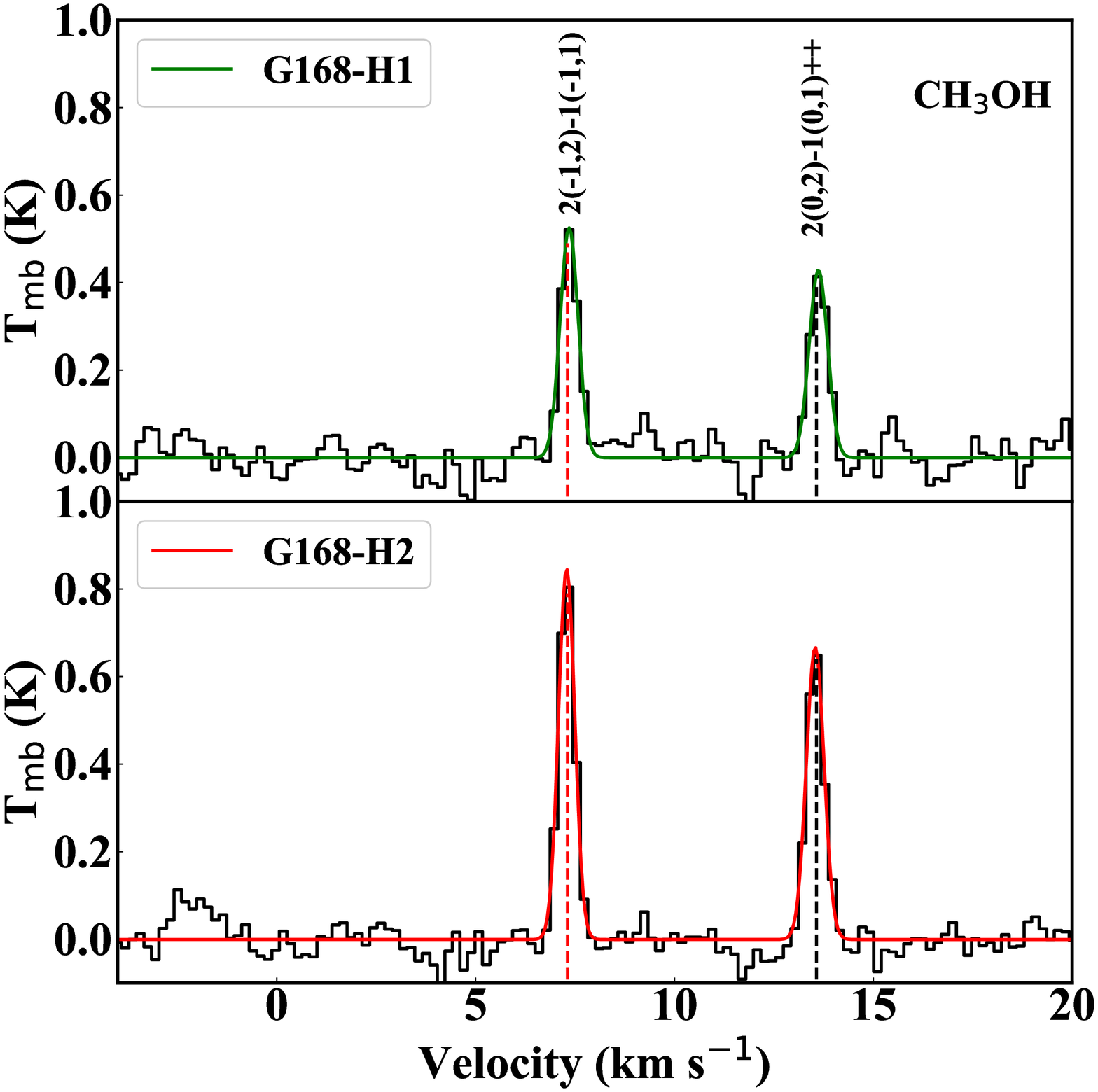}{0.3\textwidth}{(b)}
		\fig{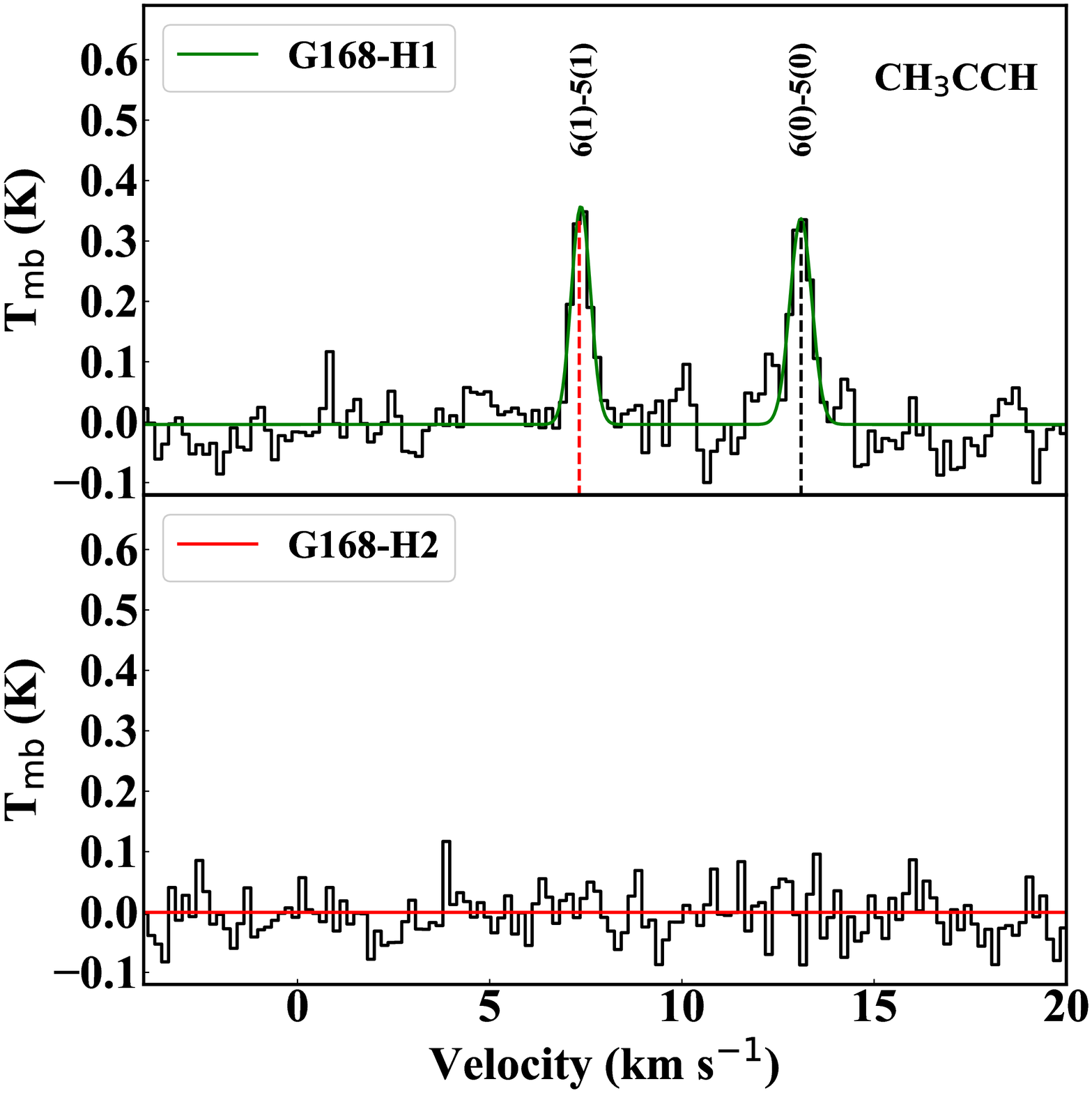}{0.3\textwidth}{(c)}
	}
	\gridline{\fig{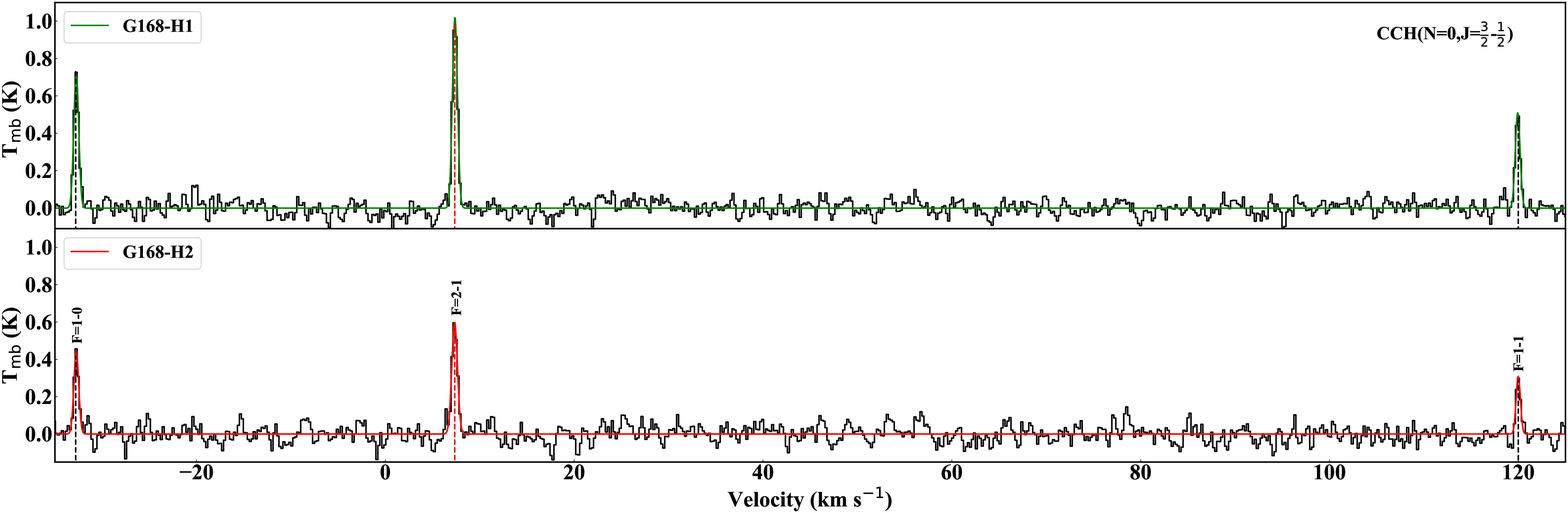}{0.99\textwidth}{(d)}}
	\caption{PMO observed transitions. These spectra were obtained by averaging over the core regions of G168-H1 and G168-H2. The black step lines in the upper and bottom panels represent the spectra of G168-H1 and G168-H2, respectively. The green and red solid lines are the Gaussian fitting results of G168-H1 and G168-H2, respectively. The  hyperfine structures are indicated by vertical dashed lines, and the red vertical dashed line denotes the velocity component corresponding to the systemic velocity. }
	\label{PMO_spec}
\end{figure*}

In Figure~\ref{PMO_spec}, we present the observed spectra averaged over the core regions of G168-H1 and G168-H2.
The line emissions of \ce{CCH} and \ce{N2H+} are stronger in G168-H1 than in G168-H2.
To the contrary, \ce{CH3OH} emission show weaker line intensity in G168-H1 than in G168-H2.
\ce{CH3CCH} was detected only in G168-H1 but not in G168-H2.
Gaussian fitting was performed to derive the line parameters of each transition, and the fitting results are summarized in
Table~\ref{Tab_base}.
The hyperfine components of $F_{1}$=1-1 and $F_{1}$=2-1 of \ce{N2H+} are blended with each other, thus, the \ce{N2H+} line parameters of T$_{\rm peak}$, FWHM, and integrated intensity cannot be obtained accurately and exhibit larger errors, as  shown in Table~\ref{Tab_base}. For blended hyperfine components of \ce{N2H+}, the rest frequency of the main component was used to determine their velocities with respect to the main component, then we used the relative velocity to estimate the velocity ranges of the blended transitions for Gaussian fitting.
From Table~\ref{Tab_base}, we found that the linewidths of all species are comparable in both G168-H1 and G168-H2 around 0.6~km s$^{-1}$. Such small linewidths may suggest that the G168-H1 and G168-H2 cores are quiescent and have not been affected by energetic gas dynamics.

\begin{figure*}
	\gridline{\fig{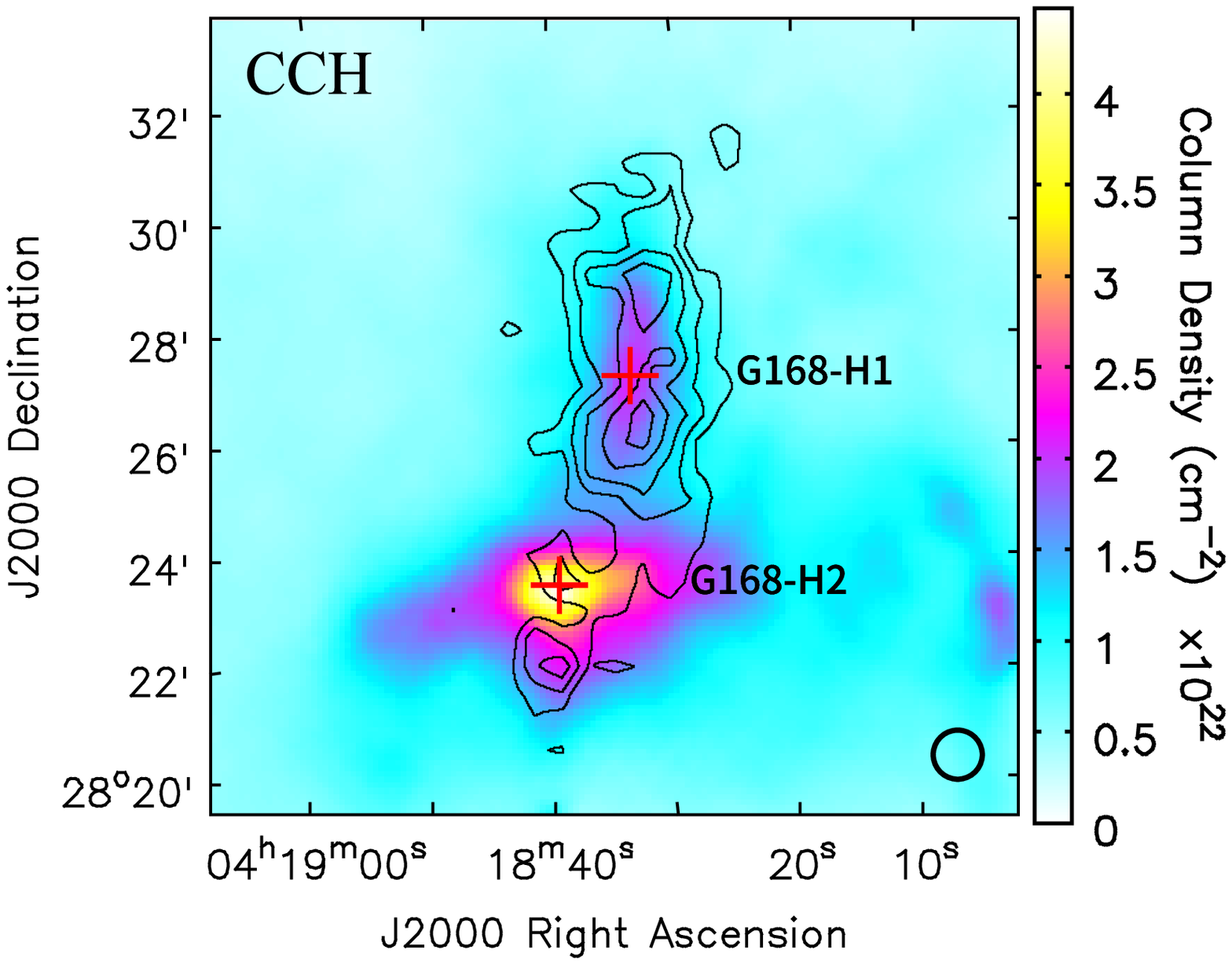}{0.45\textwidth}{(a)}
		\fig{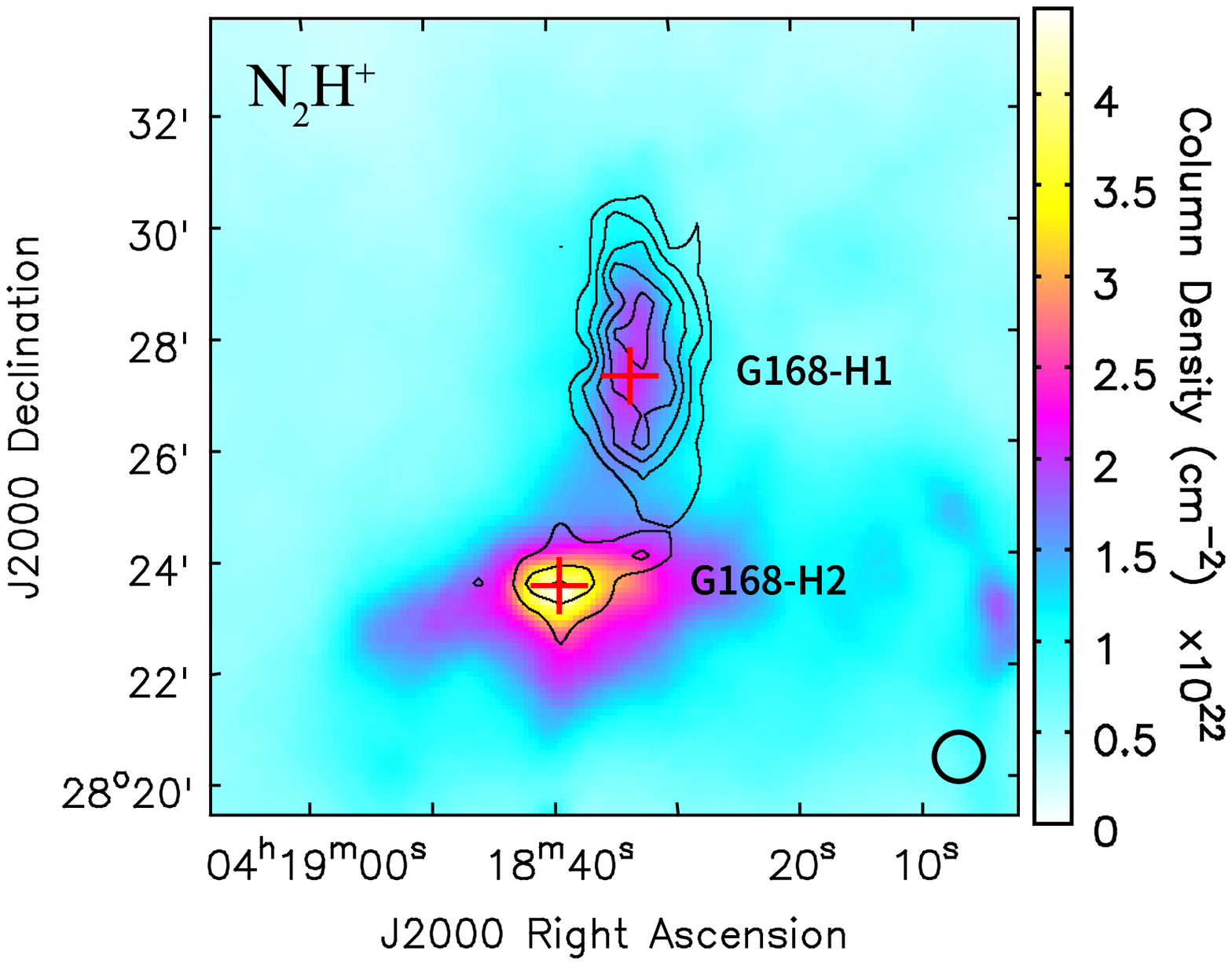}{0.45\textwidth}{(b)}
	}
	\gridline{\fig{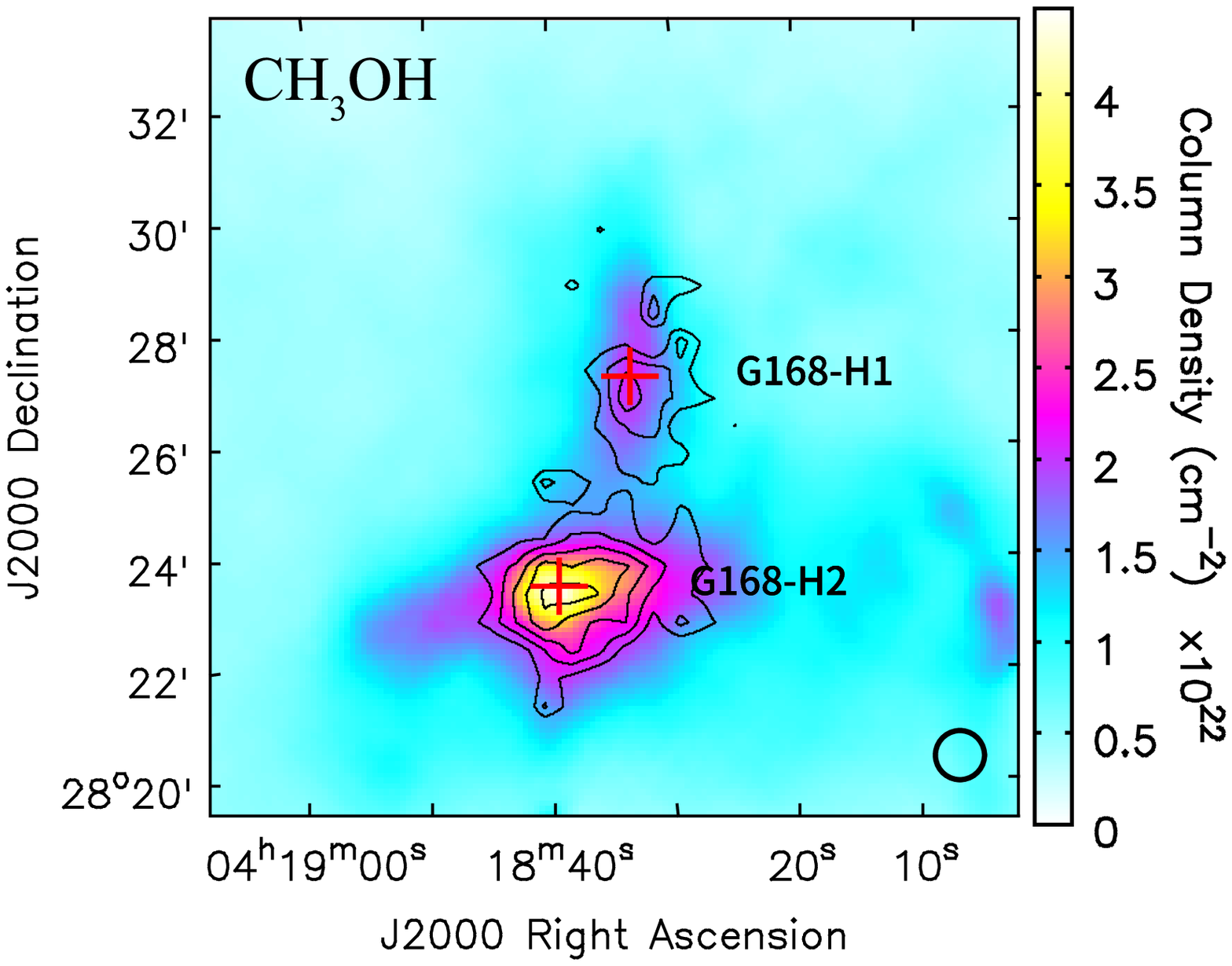}{0.45\textwidth}{(c)}
		\fig{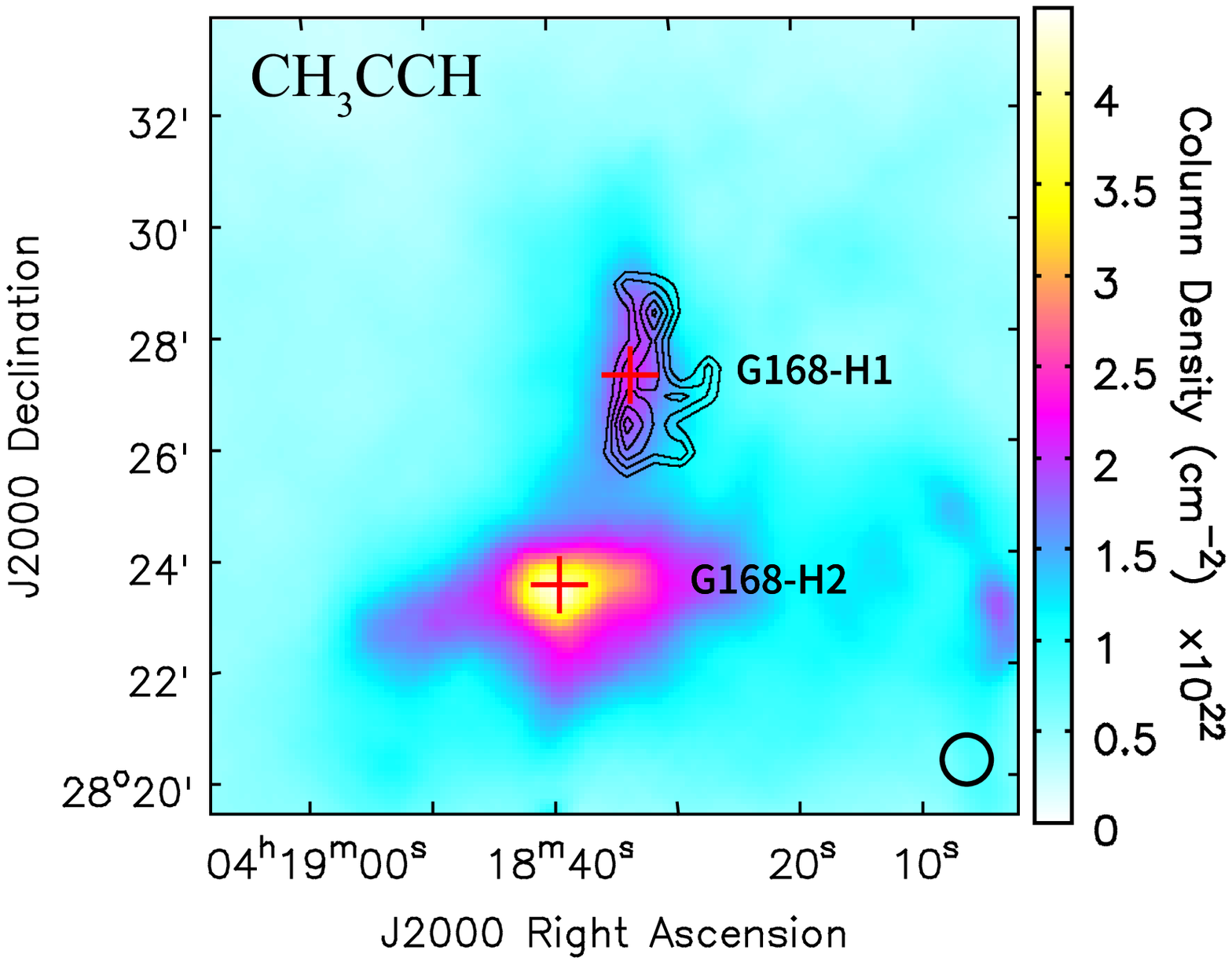}{0.45\textwidth}{(d)}
	}
	\caption{Integrated intensity maps of CCH(1-0), \ce{N2H+}(1-0), \ce{CH3OH}(2-1), and \ce{CH3CCH}(6-5) are shown in black contours in panels (a)--(d), respectively. The line name is shown in the upper left corner of each panel. The contour levels are from 30\% to 90\%, with steps of 15\% of peak integrated intensities. The maximum values of integrated intensities of \ce{CCH}, \ce{N2H+}, \ce{CH3OH}, and \ce{CH3CCH} are 1.0, 1.8, 0.7, and 0.3 K km s$^{-1}$, respectively.  The color scale in each panel represents the \emph{Herschel} \ce{H2} column density distributions from \citet{Tang18}. The red crosses denote the peak positions of the  dense cores of G168-H1 and G168-H2. The beam size of the PMO observations is shown in the bottom-right corner of each panel.
	}
	\label{PMO&Herschel}
\end{figure*}

Figure~\ref{PMO&Herschel} depicts the integrated intensity map of \ce{CCH}, \ce{N2H+}, \ce{CH3OH}, and \ce{CH3CCH} overlaid on \emph{Herschel} column density color maps \citep{Tang18}.
Note that all hyperfine components of the \ce{N2H+} $J$=1-0 transition are adopted for deriving the integrated intensity map.
In Figure~\ref{PMO&Herschel}, the gas distributions of all molecular species show north-south elongated source structures in the G168-H1 core.
Toward the G168-H2 core, the distribution of \ce{CCH} is completely different from the \emph{Herschel} H$_{2}$ column density distribution.
The distribution of \ce{N2H+} in G168-H2 is concentrated only toward the central region of the \emph{Herschel} H$_{2}$ column density map.
The \ce{CH3OH} emission appears well coupled to the \emph{Herschel} \ce{H2} column density map in both G168-H1 and G168-H2 but shows stronger emission in G168-H2 than in G168-H1.
\ce{CH3CCH} gas appears more compact than the other species in G168-H1.
There is, however, no \ce{CH3CCH} emission detected in the G168-H2 core.

\begin{deluxetable*}{cccccccccccccccccc}
	\tabletypesize{\tiny}  \tablecolumns{17}
	\tablewidth{0pc} \setlength{\tabcolsep}{0.04in}
	\rotate
	\tablecaption{PMO and TMRT Observed Lines} \tablehead{
		\colhead{Observation} &\colhead{Molecule} &\colhead{Transition} &\colhead{Frequency}  &\colhead{$S_{\rm ij}\mu^2$} &$E_{\rm u}$ &\multicolumn{2}{c}{$V_{\rm lsr}$} & &\multicolumn{2}{c}{T$_{\rm peak}$} & &\multicolumn{2}{c}{FWHM} & &\multicolumn{2}{c}{Integrated Intensity} &\colhead{Remark}\\
		\colhead{}            &\colhead{}         &\colhead{}           &\colhead{(\rm MHz)}  &          &\colhead{(K)} &\multicolumn{2}{c}{(\rm km s$^{-1}$)} &  &\multicolumn{2}{c}{(\rm K)}     &   &\multicolumn{2}{c}{(\rm km s$^{-1}$)} & & \multicolumn{2}{c}{(\rm K km s$^{-1}$)}  & \\
		\cline{7-8} \cline{10-11} \cline{13-14} \cline{16-17}
		& & & &  & &\colhead{H1} &\colhead{H2} & &\colhead{H1} & \colhead{H2}  & &\colhead{H1} & \colhead{H2} & &\colhead{H1} & \colhead{H2} &
	}
	\startdata
	TMRT &H$_{2}$CO         &2(1,1)-2(1,2),$F$=3-3                                               &14488.480   &13.58     &22.62 &7.35($\pm$0.01)   &7.32($\pm$0.01)   & &-0.61($\pm$0.05) &-0.50($\pm$0.04)  & &0.68($\pm$0.03) &0.61($\pm$0.04) & &0.44($\pm$0.02) &0.32($\pm$0.01) &$\dagger$        \\
	&c-C$_{3}$H        &1(1,0)-1(1,1),$J$=$\frac{1}{2}$-$\frac{3}{2}$,F=1-2  &14767.700    &2.83      &0.71     &7.74($\pm$0.05)   &7.75($\pm$0.03)   & &0.14($\pm$0.13)  &0.15($\pm$0.07)    & &0.47($\pm$0.33) &0.21($\pm$0.06) & &0.07($\pm$0.03) &0.03($\pm$0.01) &$\dagger$        \\
	&                          &1(1,0)-1(1,1),$J$=$\frac{3}{2}$-$\frac{3}{2}$,F=2-2 &14893.050  &11.12       &0.71    &7.34($\pm$0.02)   &7.27($\pm$0.01)   & &0.28($\pm$0.05)  &0.33($\pm$0.03)   & &0.42($\pm$0.05) &0.32($\pm$0.01) & &0.13($\pm$0.01) &0.11($\pm$0.01) &$\dagger$ $\ast$ \\
	&HC$_{5}$N         &$J$=5-4,F=4-3                                                       &13313.261    &24.30    &1.92    &8.95($\pm$0.01)   &8.99($\pm$0.03)   & &0.33($\pm$0.03)  &0.12($\pm$0.05)   & &0.41($\pm$0.02) &0.28($\pm$0.06) & &0.15($\pm$0.01) &0.03($\pm$0.01) &$\ast$           \\
	&                          &$J$=5-4,$F$=5-4                                                   &13313.312    &30.00    &1.92    &7.71($\pm$0.02)   &7.79($\pm$0.01)   & &0.41($\pm$0.04)  &0.16($\pm$0.37)  &  &0.54($\pm$0.04) &0.13($\pm$0.33) & &0.24($\pm$0.01) &0.02($\pm$0.01) &                 \\
	&                          &$J$=5-4$,F$=6-5                                                   &13313.334   &36.93    &1.92    &7.29($\pm$0.01)   &7.29($\pm$0.01)   & &0.47($\pm$0.05)  &0.20($\pm$0.15)   & &0.29($\pm$0.02) &0.16($\pm$0.11) & &0.15($\pm$0.01) &0.03($\pm$0.01) &$\dagger$        \\
	&HC$_{7}$N         &$J$=11-10                                                               &12408.003   &255.56  &3.57    &7.28($\pm$0.02)   &...               & &0.13($\pm$0.09)  &...               & &0.51($\pm$0.04) &...             & &0.07($\pm$0.05) &...             &$\dagger$        \\
	&                          &$J$=12-11                                                               &13535.998   &278.82  &4.22    &7.29($\pm$0.01)   &...               & &0.34($\pm$0.04)  &...               & &0.46($\pm$0.03) &...             & &0.17($\pm$0.01) &...             &$\dagger$ $\ast$ \\
	&                          &$J$=13-12                                                              &14663.993   &302.05  &4.93    &7.27($\pm$0.02)   &...               & &0.38($\pm$0.07)  &...               & &0.42($\pm$0.04) &...             & &0.17($\pm$0.02) &...             &$\dagger$        \\
	&SO                      &1(2)-1(1)                                                                  &13043.346   &3.45      &15.81    &...               &7.41($\pm$0.02)   & &...              &0.19($\pm$0.05)   & &...             &0.35($\pm$0.06) & &...             &0.07($\pm$0.01) &$\dagger$ $\ast$ \\
	\cline{1-18}
	PMO  &CCH                    &$N$=1-0,$J$=$\frac{3}{2}$-$\frac{1}{2}$,F=1-1     &87284.156   &0.15     &4.19   &119.10($\pm$0.01) &119.16($\pm$0.04) & &0.51($\pm$0.06) &0.31($\pm$0.06)    & &0.60($\pm$0.04) &0.54($\pm$0.07) & &0.33($\pm$0.02) &0.18($\pm$0.02) &                 \\
	&                          &$N$=1-0,$J$=$\frac{3}{2}$-$\frac{1}{2}$,F=2-1    &87316.925   &1.43     &4.19   &7.35($\pm$0.01)   &7.35($\pm$0.02)   & &1.02($\pm$0.05) &0.60($\pm$0.07)    & &0.64($\pm$0.02) &0.71($\pm$0.05) & &0.69($\pm$0.02) &0.45($\pm$0.03) &$\dagger$ $\ast$ \\
	&                          &$N$=1-0,$J$=$\frac{3}{2}$-$\frac{1}{2}$,F=1-0    &87328.624   &0.91    &4.19   &-33.18($\pm$0.01) &-33.17($\pm$0.03) & &0.72($\pm$0.06) &0.44($\pm$0.07)    & &0.61($\pm$0.03) &0.60($\pm$0.07) & &0.47($\pm$0.02) &0.28($\pm$0.02) &                 \\
	&N$_{2}$H$^{+}$ &$J$=1-0,$F_{1}$=1-1,F=0-1                                      &93171.621    &37.25   &4.47   &14.26($\pm$0.17)  &14.28($\pm$0.13)  & &0.22($\pm$0.31) &0.17($\pm$0.19)    & &0.48($\pm$0.43) &0.43($\pm$0.28) & &0.11($\pm$0.09) &0.08($\pm$0.05) &                 \\
	&                          &$J$=1-0,$F_{1}$=1-1,$F$=2-2                                 &93171.917     &...        &...       &13.31($\pm$0.53)  &13.33($\pm$0.47)  & &0.61($\pm$1.38) &0.32($\pm$0.21)    & &0.63($\pm$0.67) &0.66($\pm$0.29) & &0.41($\pm$0.65) &0.23($\pm$0.07) &                 \\
	&                          &$J$=1-0,$F_{1}$=1-1,$F$=1-0                                  &93172.053   &...        &...       &12.85($\pm$0.52)  &12.87($\pm$0.29)  & &0.37($\pm$0.92) &0.22($\pm$0.18)    & &0.48($\pm$0.61) &0.44($\pm$0.26) & &0.19($\pm$0.31) &0.10($\pm$0.04) &                 \\
	&                          &$J$=1-0,$F_{1}$=2-1,$F$=2-1                                 &93173.480   &62.09  &4.47    &8.325($\pm$0.06)  &8.31($\pm$0.06)   & &0.58($\pm$0.24) &0.33($\pm$0.17)    & &0.68($\pm$0.18) &0.62($\pm$0.18) & &0.42($\pm$0.09) &0.21($\pm$0.07) &                 \\
	&                          &$J$=1-0,$F_{1}$=2-1,$F$=3-2                                &93173.777     &...       &...        &7.38($\pm$0.10)   &7.38($\pm$0.08)   & &0.81($\pm$0.66) &0.47($\pm$0.33)    & &0.54($\pm$0.21) &0.47($\pm$0.19) & &0.46($\pm$0.27) &0.23($\pm$0.10) &$\dagger$        \\
	&                          &$J$=1-0,$F_{1}$=2-1,$F$=1-1                                 &93173.967    &...        &...        &6.81($\pm$0.29)   &6.80($\pm$0.15)   & &0.37($\pm$0.59) &0.25($\pm$0.45)    & &0.68($\pm$0.56) &0.53($\pm$0.41) & &0.28($\pm$0.27) &0.14($\pm$0.19) &                 \\
	&                          &$J$=1-0,$F_{1}$=0-1,$F$=1-2                                &93176.265    &12.42  &4.47     &-0.66($\pm$0.06)  &-0.68($\pm$0.05)  & &0.45($\pm$0.30) &0.31($\pm$0.18)    & &0.58($\pm$0.25) &0.46($\pm$0.15) & &0.28($\pm$0.10) &0.16($\pm$0.05) &$\ast$           \\
	&CH$_{3}$OH       &2(-1,2)-1(-1,1)                                                        &96739.362    &1.21     &12.54   &13.53($\pm$0.02)  &13.44($\pm$0.01)  & &0.43($\pm$0.06) &0.68($\pm$0.06)    & &0.54($\pm$0.05) &0.54($\pm$0.03) & &0.25($\pm$0.02) &0.39($\pm$0.02) &                 \\
	&                          &2(0,2)-1(0,1)++                                                      &96741.375     &1.62    &6.97     &7.26($\pm$0.02)   &7.20($\pm$0.01)   & &0.53($\pm$0.06) &0.85($\pm$0.06)    & &0.53($\pm$0.04) &0.48($\pm$0.02) & &0.30($\pm$0.02) &0.44($\pm$0.02) &$\dagger$ $\ast$ \\
	&CH$_{3}$CCH    &6(1)-5(1)                                                                &102546.024  &2.18     &24.45   &13.18($\pm$0.03)  &...               & &0.34($\pm$0.12) &...                & &0.68($\pm$0.09) &...             & &0.25($\pm$0.07) &...             &                 \\
	&                          &6(0)-5(0)                                                              &102547.984   &2.25    &17.23    &7.46($\pm$0.03)   &...               & &0.36($\pm$0.06) &...                & &0.59($\pm$0.06) &...             & &0.23($\pm$0.02) &...             &$\dagger$ $\ast$ \\
	\enddata
	\label{Tab_base}
	\tablecomments{($\dagger$):The velocity components consistent with their systemic velocities are marked with ``$\dagger$''. ($\ast$):The transitions marked with ``$\ast$'', given their high signal-to-noise ratios, were adopted for calculating the molecular column densities and abundances.}
\end{deluxetable*}

\subsection{Column density and abundance}
Under the assumption of local thermodynamic equilibrium (LTE) , the molecular column density can be calculated using the following equation \citep{Mangum15}:
\begin{equation}
\it{N = \frac{3 h}{8 \pi^{3} S \mu^{2}} \frac{Q(T_{\rm ex})}{g_{u}} \frac{e^{\left(\frac{E_{\rm u}}{k T_{\rm ex}}\right)}}{e^{(\frac{h \nu}{k T_{\rm ex}})} - 1} \int \tau dV },
\label{N}
\end{equation}
where $\nu$ is the rest frequency of the specific transition, $T_{\rm ex}$ is the excitation temperature, $g_{\rm u}$ is the level degeneracy, $h$ is the Planck constant, $S \mu^{2}$ is the product of the total torsion rotational line strength and the square of the electric dipole moment, $Q(T_{\rm ex})$ is the partition function at a temperature of $T_{\rm ex}$, and $E_{\rm u}$ is the upper level energy.
The numerical values of these parameters are taken from Splatalogue$\footnote{http://www.cv.nrao.edu/php/splat/}$.
The GILDAS package provides HFS fitting to derive the optical depth ($\tau$) of the molecules with hyperfine components.
The derived optical depths of \ce{N2H+}($J=1-0, F_{1}=0-1, F=1-2$) are 0.67($\pm$0.14) and 2.20($\pm$0.35) for G168-H1 and G168-H2, respectively.
The relationship between brightness temperature ($T_{\rm b}$), optical depth ($\tau$), and excitation temperature ($T_{\rm ex}$)  is as follows:
\begin{equation}
\it{T_{\rm b} =f\left[J(T_{\rm ex}) - J(T_{\rm bg})\right]\times(1 -e^{-\tau})},
\end{equation}
\begin{equation}
\it{J(T_{\rm ex}) = \frac{\frac{h \nu}{k}}{e^{(h \nu / k T_{\rm ex})}- 1}},
\end{equation}
where $T_{\rm bg}$ = 2.73 K is the background temperature. $f$ = $\frac{\Omega_{\rm source}}{\Omega_{\rm source} + \Omega_{\rm beam}}$ = 0.8 is filling factor, and $\Omega_{\rm source}$ and $\Omega_{\rm beam}$ are the solid angles of source and beam, respectively. The brightness temperature ($T_{\rm b}$) of \ce{N2H+} is obtained from the Gaussian fitting as listed in Table~\ref{Tab_base} .
Thus, the excitation temperatures of \ce{N2H+} in G168-H1 and G168-H2 are estimated to be 3.92($\pm$0.77) K and 3.20($\pm$0.26) K, respectively.
We have also estimated the excitation temperature of \ce{N2H+} by using the non-LTE calculation tool of Radex \citep{Van07}, which gives $T_{\rm ex}$ of 3.9 and 3.4 K for G168-H1 and G168-H2, respectively.
Similar excitation temperatures derived from the two methods suggest that the estimated excitation temperatures are reliable. Therefore, we adopted the excitation temperatures of \ce{N2H+} derived from the hyperfine components to estimate the molecular column densities for all the observed species.
Note that not all molecules have the same excitation temperature, we use $T_{\rm ex}$ of \ce{N2H+} to estimate the column densities because of the lack of sufficient  transitions to derive temperatures for the other species.
The column densities of the detected species are listed in Table~\ref{Tab_cal}.
The uncertainties of the derived molecular column densities mainly result from the following:
(1) 20\% and 7\% uncertainties from TMRT's injecting periodic noise amplitude calibration \citep{Li17} and PMO's chopper-wheel method amplitude calibration  \citep{Ulich76}, respectively;
(2) the uncertainty from Gaussian fitting toward the spectrum;
and (3) the uncertainty from excitation temperature ($T_{\rm ex}$), which contributes 1\% -- 10\% uncertainties to molecular column densities of different species.
The uncertainty of the derived column density ranges from 13\% to 40\%.
\ce{H2CO} is seen in absorption, which is only a ``foreground'' absorption layer of \ce{H2CO} and has been excluded from further analyses.

\citet{Tang18} have derived \ce{H2} column densities of 1.40($\pm$0.21)$\times$10$^{22}$ and 2.6($\pm$0.62)$\times$10$^{22}$ cm$^{-2}$ for G168-H1 and G168-H2, respectively.
Taking these \ce{H2} column densities into consideration, molecular fractional abundances (\emph{X} = \emph{N$_{\rm x}$}/\emph{N$_{\rm H_{2}}$}) of the observed molecules can be obtained. We list the abundances of all molecules in Table~\ref{Tab_cal}.
The upper limits of the column density and fractional abundance for the nondetected lines were also estimated by using three times rms noise (3$\sigma$) and the mean FWHM of the spectra.

\begin{deluxetable*}{cccccccccc}
\tabletypesize{\scriptsize}
\tablewidth{0pt}
\tablecaption{Derived Parameters}
\tablehead{
\colhead{Observation} &\colhead{Molecule} &\multicolumn{2}{c}{$Q(T_{\rm ex})^{\ast}$} &\colhead{} &\multicolumn{2}{c}{Column Density ($N$)} &\colhead{} &\multicolumn{2}{c}{Abundance ($X$)}  \\
\colhead{}            &\colhead{}         &\multicolumn{2}{c}{}              &\colhead{} &\multicolumn{2}{c}{(\rm cm$^{-2}$)}      &\colhead{} &\multicolumn{2}{c}{}         \\
\cline{3-4} \cline{6-7} \cline{9-10}
\colhead{}            &\colhead{}         &\colhead{H1} &\colhead{H2}        &\colhead{} &\colhead{H1} & \colhead{H2}              &\colhead{} &\colhead{H1} &\colhead{H2}
}
\startdata
TMRT     &c-C$_{3}$H      &38.08  &36.38  & &2.14($\pm$1.13)$\times$10$^{13}$  &1.72($\pm$0.68)$\times$10$^{13}$      & &1.53($\pm$0.84)$\times$10$^{-9}$      &6.98($\pm$3.25)$\times$10$^{-10}$   \\
         &HC$_{5}$N       &107.09 &97.65  & &4.16($\pm$1.31)$\times$10$^{12}$  &1.73($\pm$0.69)$\times$10$^{12}$      & &2.97($\pm$1.03)$\times$10$^{-10}$     &7.03($\pm$3.28)$\times$10$^{-11}$   \\
         &HC$_{7}$N       &420.27 &328.11 & &1.25($\pm$0.39)$\times$10$^{12}$  &$\textless$4.75$\times$10$^{11}$      & &8.95($\pm$3.09)$\times$10$^{-11}$     &$\textless$1.93$\times$10$^{-11}$  \\
         &SO              &3.344  &1.63   & &$\textless$2.32$\times$10$^{13}$  &1.37($\pm$0.35)$\times$10$^{14}$      & &$\textless$1.66$\times$10$^{-9}$      &5.58($\pm$1.98)$\times$10$^{-9}$   \\
\cline{1-10}
PMO  &CCH                 &9.48   &8.05   & &1.19($\pm$0.29)$\times$10$^{13}$  &1.56($\pm$0.20)$\times$10$^{13}$      & &8.52($\pm$2.43)$\times$10$^{-10}$     &6.34($\pm$1.74)$\times$10$^{-10}$   \\
     &N$_{2}$H$^{+}$      &77.62  &76.67  & &7.47($\pm$3.23)$\times$10$^{12}$  &1.01($\pm$0.33)$\times$10$^{13}$      & &5.34($\pm$2.43)$\times$10$^{-10}$     &4.08($\pm$1.68)$\times$10$^{-10}$   \\
     &CH$_{3}$OH          &11.21  &9.91   & &9.55($\pm$2.42)$\times$10$^{12}$  &3.40($\pm$0.40)$\times$10$^{13}$      & &6.82($\pm$1.99)$\times$10$^{-10}$     &1.38($\pm$0.37)$\times$10$^{-9}$   \\
     &CH$_{3}$CCH         &17.61  &15.15  & &4.03($\pm$1.05)$\times$10$^{13}$  &$\textless$3.48$\times$10$^{13}$      & &2.88($\pm$0.85)$\times$10$^{-9}$      &$\textless$1.42$\times$10$^{-9}$  \\
\enddata
\label{Tab_cal}
\tablecomments{$Q(T_{\rm ex})$ is the partition function corresponding to excitation temperatures of 3.92 K and 3.20 K for the G168-H1 and G168-H2 cores, respectively. For the nondetected lines, we derived their upper limits based on three times rms noise (3$\sigma$) and marked them with ``$\textless$''.}
\end{deluxetable*}

\begin{figure}
\plotone{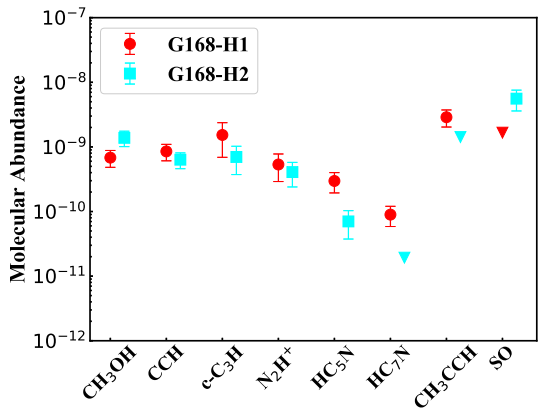}
\caption{Molecular abundances of observed species (points with error bars) in G168-H1 (red) and G168-H2 (cyan). \ce{CH3CCH} and \ce{HC7N} were detected only in the G168-H1 core, and SO was detected only in the G168-H2 core. The down-pointing triangles indicate the upper limits derived from observations.}
\label{fig_abun}
\end{figure}

\section{Discussion}
\subsection{Observed chemical differences between G168-H1 and G168-H2}
\label{chemical differences}
Figure~\ref{fig_abun} shows the molecular abundances of all observed species.
According to Figure~\ref{fig_abun}, most molecular abundances are lower in G168-H2 than those in G168-H1.
It is clear that the carbon-chain molecules (especially for \ce{HC5N}, \ce{HC7N}, and \ce{CH3CCH}) are more abundant in G168-H1 than in G168-H2.
Meanwhile, the SO line exhibits an opposite trend compared with other species, showing high abundance in the G168-H2 core and nondetection in the G168-H1 core.
 G168-H1 and G168-H2 present disparate features in their molecular spectra, gas distributions, and molecular abundances, strongly implying that the two cores likely have different chemical properties.

The carbon-chain molecules CCH and \ce{HC5N} show stronger line emissions in G168-H1 than those in G168-H2, although the G168-H2 continuum emission is stronger than that of G168-H1.
In particular, the \ce{HC7N} and \ce{CH3CCH} lines were totally absent in the G168-H2 core.
Mapping observations with the PMO also confirm the trend that the carbon-chain molecules are denser in the G168-H1 core than those in the G168-H2 core.
From Table~\ref{Tab_cal}, one can see that carbon-chain molecules have higher abundances in G168-H1 than those in G168-H2.
Carbon-chain molecules may trace the early stages of dense cores, and they have been found to be deficient in the evolved central parts of dense cores \citep{Ohashi99,Tafalla98,Aikawa01}.
As we mentioned before, G168-H2 seems to be more evolved than G168-H1.
The observed differences in the carbon-chain molecular abundances between G168-H1 and G168-H2 are probably due to the fact that they are at different evolutionary stages.

As shown in panel (h) of Figure~\ref{TMRT_spec}, the SO line emission was detected only in the G168-H2 core.
Sulfur-bearing molecules have been found to be depleted in starless and prestellar cores \citep{Tafalla06,Agundez13,Vastel18}.
Studies of sulfur-bearing molecules suggested that sulfur is initially in the form of \ce{S+}, which can efficiently produce a S atom by electronic recombination mechanism. Thus, the main reservoirs of sulfur could be a S atom in the gas phase during $10^{3}\sim10^{5}$ yr in the cold dark clouds. \citep{Bockelee00,Vidal17,Lass19}.
Later, the S atom is consumed by the gas-phase reaction with \ce{H3+}, \ce{CH3}, \ce{OH}, and \ce{O2} to efficiently produce SO \citep{Vidal17}. As G168-H2 is more evolved and denser than G168-H1, there may be sufficient time to produce abundant SO in G168-H2.
A possible reason for the nondetection of SO in G168-H1 is that G168-H1 is less evolved; thus, SO may not reach detectable abundance in G168-H1 in a short time.

Figure~\ref{PMO_spec} shows stronger \ce{CH3OH} line emission in G168-H2 than that in G168-H1.
Figure~\ref{PMO&Herschel} also shows denser gas distribution in G168-H2 than that in G168-H1.
\ce{CH3OH} has been observed in cold dark clouds of TMC-1 CP, L1498, L1517B, and L1544, suggesting that \ce{CH3OH} is produced on dust grain and then liberated into the gas phase through nonthermal desorption processes in cold dark clouds \citep{Pratap97,Takakuwa00,Tafalla06,Vastel14}.
As shown in panel (c) of Figure~\ref{PMO&Herschel}, the \ce{CH3OH} gas distribution is well coupled with dust in both G168-H1 and G168-H2. This may indicate that \ce{CH3OH} originates from grain-surface chemistry \citep{Taquet2012}.

Previous observations suggested that \ce{N2H+} is mainly concentrated in the center of the core and is more abundant in more evolved dense cores \citep{Tafalla98,Crapsi05,SakaiT08}.
Therefore, \ce{N2H+} has been generally used as an evolutionary tracer of dense cores.
As stated before, G168-H2 is more evolved than G168-H1. Thus, we would expect that the abundance of \ce{N2H+} in G168-H2 is higher than that in G168-H1.
However, the G168-H1 and G168-H2 cores have similar \ce{N2H+} abundances, which conflicts with the previous studies  with higher \ce{N2H+} abundances in the evolved cores.
Additionally, from panel (b) of Figure~\ref{PMO&Herschel}, one can see that distribution of \ce{N2H+} completely covers the dust emission in G168-H1, whereas \ce{N2H+} emission is concentrated only in a very small region around the peak of the dust in G168-H2.
\citet{Lippok13} studied seven isolated starless cores by observations and simulations, suggesting that \ce{N2H+} is depleted in the center of the cores with hydrogen volume density $\ga$10$^{5}$ cm$^{-3}$ and column density $\textgreater$10$^{22}$ cm$^{-2}$. As listed in Table~\ref{Tab_para}, the central \ce{H2} volume density and column density of G168-H2 is 1.77($\pm$0.13)$\times$10$^{5}$ cm$^{-3}$ and 4.50($\pm$0.01)$\times$10$^{22}$ cm$^{-2}$, respectively. Thus, we speculate that the exceptionally weak emission of \ce{N2H+} in G168-H2 may be caused by \ce{N2H+} depletion.
\ce{N2H+} is produced by \ce{N2} and destroyed by CO in cold dark clouds by the gas reactions of \ce{N2} + \ce{H3+} $\rightarrow$ \ce{N2H+} + \ce{H2} and \ce{N2H+} + CO $\rightarrow$ \ce{HCO+} + \ce{N2}, respectively \citep{Aikawa01,Bergin02}.
When compared with other prestellar cores and evolved cloud cores, the lower abundance of \ce{N2H+} in prestellar core G168-H2 is probably caused by an increase in CO or \ce{N2} depletion.
\citet{Lippok13} studied the correlation between C$^{18}$O and \ce{N2H+} that the \ce{N2H+} and C$^{18}$O abundances are anticorrelated when the C$^{18}$O abundance exceeds 10$^{-8}$. As shown in Table~\ref{Tab_para}, the C$^{18}$O abundances of G168-H1 and G168-H2 are 1.21($\pm$0.29)$\times$10$^{-7}$ and 0.91($\pm$0.29)$\times$10$^{-7}$, respectively.
Therefore, although destruction by CO is not negligible, we suggest that the \ce{N2H+} depletion in the G168-H2 core is dominated by \ce{N2} depletion rather than destruction by CO.

\subsection{Comparison with other objects}
\label{comparison}
The fractional abundances of carbon-chain molecules in cold dark clouds are approximately $(0.1\sim4.9)\times10^{-9}$ for \ce{HC5N}, $(0.2\sim7.4)\times10^{-10}$ for \ce{HC7N}, and $(0.1\sim8.1)\times10^{-9}$ for \ce{CH3CCH} \citep{Benson83,SakaiN08,SakaiN09,SakaiN10,Friesen13,Vasyunina14,Burkhardt18,Mendoza18,Seo19,Wu19}.
Our observations gave  abundances of \ce{HC5N}, \ce{HC7N}, and \ce{CH3CCH} of 2.97($\pm$1.03)$\times$10$^{-10}$, 8.95($\pm$3.09)$\times$10$^{-11}$, and 2.88($\pm$0.85)$\times$10$^{-9}$ in G168-H1, respectively.
These values, though at the lower end, are comparable to those of cold dark clouds.
However, \ce{HC7N} and \ce{CH3CCH} were not detected in G168-H2.
Our observations measured an \ce{HC5N} abundance of only 7.03($\pm$3.28)$\times$10$^{-11}$ in G168-H2,
which is one order of magnitude lower than those of cold dark clouds.
Our results suggest that the starless core G168-H1 may exhibit properties of cold dark clouds, whereas the prestellar core G168-H2 has lower carbon-chain molecular abundances than general cold dark clouds do.

Past studies reported an \ce{N2H+} abundance of $(0.7\sim9)\times10^{-10}$ in cold dark clouds \citep{Benson98,Bergin02,Caselli02,Friesen10,Taniguchi19}.
The derived \ce{N2H+} abundances of G168-H1 and G168-H2 are 5.34($\pm$2.43)$\times$10$^{-10}$ and 4.08($\pm$1.68)$\times$10$^{-10}$, respectively, which are comparable with previous measurements toward cold dark clouds.

Previous observations reported a  \ce{CH3OH} abundance of $(0.52\sim47)\times10^{-9}$ in cold dark clouds \citep{Gomez11,Pratap97,Requena07,Soma15,Tafalla06,Vasyunina14,Vastel14}.
The estimated abundances of \ce{CH3OH} are
6.82($\pm$1.99)$\times$10$^{-10}$ and 1.38($\pm$0.37)$\times$10$^{-9}$ in G168-H1 and G168-H2, respectively, which coincide with those in cold dark clouds. Unlike the abundance differences for the other molecules, the \ce{CH3OH} abundances are lower in G168-H1 than in G168-H2, indicating that \ce{CH3OH} shows different chemical properties from carbon-chain and nitrogen-bearing molecules---its abundance increases with core evolution and does not exhibit depletion in PGCC G168.

\subsection{Chemical model}
\label{chemical model}
To verify the chemical property of the observed molecules in G168-H1 and G168-H2, gas-grain chemical model calculations are highlighted in this section.

We adopt the gas-grain chemical reaction network from \citet{Semenov10}. The gas phase and solid phase are linked through accretion and desorption, which includes both thermal and cosmic-ray-induced desorption processes.
In comparison with previous gas-grain models with single-size dust grains \citep[e.g. 0.1\,$\mu$m in models of][]{Semenov10}, the MRN dust size distribution for silicate grains \citep{Mathis77} is adopted to sample dust grains in the range of 0.005--0.25\,$\mu$m in our model. Thus, the size-dependent cosmic-ray desorption proposed by \citet{Iqbal18} is used as an effective nonthermal mechanism (see more details in Appendix~\ref{physical parameters}), which alters the thermal diffusion of species on dust surface and the desorption. Thus, diffusion energy $E_{\rm diff}$ of species controlling the surface reactions becomes more important, which is usually estimated through binding energy $E_{\rm bind}$ with a given diffusion-to-binding energy ratio ($R_{\rm db}$) in the range of 0.3--0.77 \citep[e.g.][]{Semenov10}. The smaller the $R_{\rm db}$ value is, the faster the thermal diffusion of species is on the dust grain surface. Therefore, $R_{\rm db}$ is also varied to be 0.3, 0.5, and 0.77 to explore the effects of chemistry, particularly for \ce{CH3OH,} which is mainly formed on dust grains and then can be released by effective desorption. The physical parameters used for G168-H1 and G168-H2 in this study are listed in Table~\ref{tab_phy}, and the determinations of them are presented in Appendix~\ref{physical parameters}.
\begin{table}
\centering
\caption{Physical Parameters Used for Chemical Models.}
\label{tab_phy}
\begin{tabular}{llllll}
\hline
Core    &   $n_{\rm H}$       & $T_{\rm gas}$ &   $T_{\rm dust}$   & $A_{\rm V}$  & $R_{\rm db}^a$\\
        &   (cm$^{-3}$)       & (K)&   (K)  & (mag) &  \\
\hline
G168-H1  &   $2\times 10^{4}$  & 10.0 &  11.9 &  10.0 & 0.3, 0.5, 0.77\\
G168-H2  &   $1\times 10^{5}$  & 10.0 &  13.2 &  10.0  & 0.3, 0.5, 0.77\\
\hline
\end{tabular}\\ [1mm]
a: $R_{\rm db}$ is the diffusion-to-binding energy ratio.
\end{table}

For the initial abundances, the low-metal set is used (as listed in Table~\ref{tab_iab}), which has been widely used for dark cloud models \citep{Acharyya17,Wakelam2008}. In addition, the GGCHEM Fortran code was used, which has been successfully benchmarked with various models in \citet{Semenov10}, and has been used to study a number of astrochemical subjects such as the effects of turbulent dust motion on interstellar chemistry \citep{Ge16a} and the chemical differentiation across dust grain sizes \citep{Ge16b}.

\begin{table}
	\centering
	\caption{Initial Abundances.}
	\label{tab_iab}
	\begin{tabular}{llllll}
		\hline
		Species   &   $n_{i}/n_{\rm H}$      & Species &   $n_i/n_{\rm H}$     \\
		\hline
		H$_2$     &   0.5                  & S$^+$    &   $8.00\times 10^{-8}$  \\
		He        &   $9.00\times 10^{-2}$ & Fe$^+$   &   $3.00\times 10^{-9}$  \\
		C$^+$     &   $1.20\times 10^{-4}$ & Na$^+$   &   $2.00\times 10^{-9}$  \\
		N         &   $7.60\times 10^{-5}$ & Mg$^+$   &   $7.00\times 10^{-9}$  \\
		O         &   $2.56\times 10^{-4}$ & Cl$^+$   &   $1.00\times 10^{-9}$  \\
		Si$^+$    &   $8.00\times 10^{-9}$ & P$^+$    &   $2.00\times 10^{-10}$ \\
		\hline
	\end{tabular}\\ [1mm]
\tablecomments{The initial abundances are the low-metal set from \citet{Acharyya17} and \citet{Wakelam2008}.}
\end{table}

To compare models with observations, we use the chi-squared method to obtain the optimal chemical age for each model by searching for the minimum chi-squared ($\chi^2$) using the eight observed molecular abundances listed in Table~\ref{Tab_cal} (see Appendix~\ref{chemical age}). Because of the limited observational data (some species only have upper limits), the optimal chemical age is difficult to be determined precisely by the chi-square method.
Therefore, we prefer to use the optimal chemical age range, suggested by the chi-square method, as (2$\sim$3)$\times 10^5$ yr and (1$\sim$2)$\times10^5$ yr for G168-H1 and G168-H2, respectively, for further discussion.
Because our physical model does not evolve with time, the optimal chemical age is the time scale required for the certain species to reach particular abundances in agreement with the observed abundances, rather than an estimate of the physical age of the real source starts from the diffuse  state  to  the  current  dense  core.
Therefore, the chemical age depends on the physical condition of the core and initial chemical abundances and does not reflect the physical age of the core \citep{Maret2013,Bergin06}.

\begin{figure*}
\centering
\includegraphics[scale=0.73]{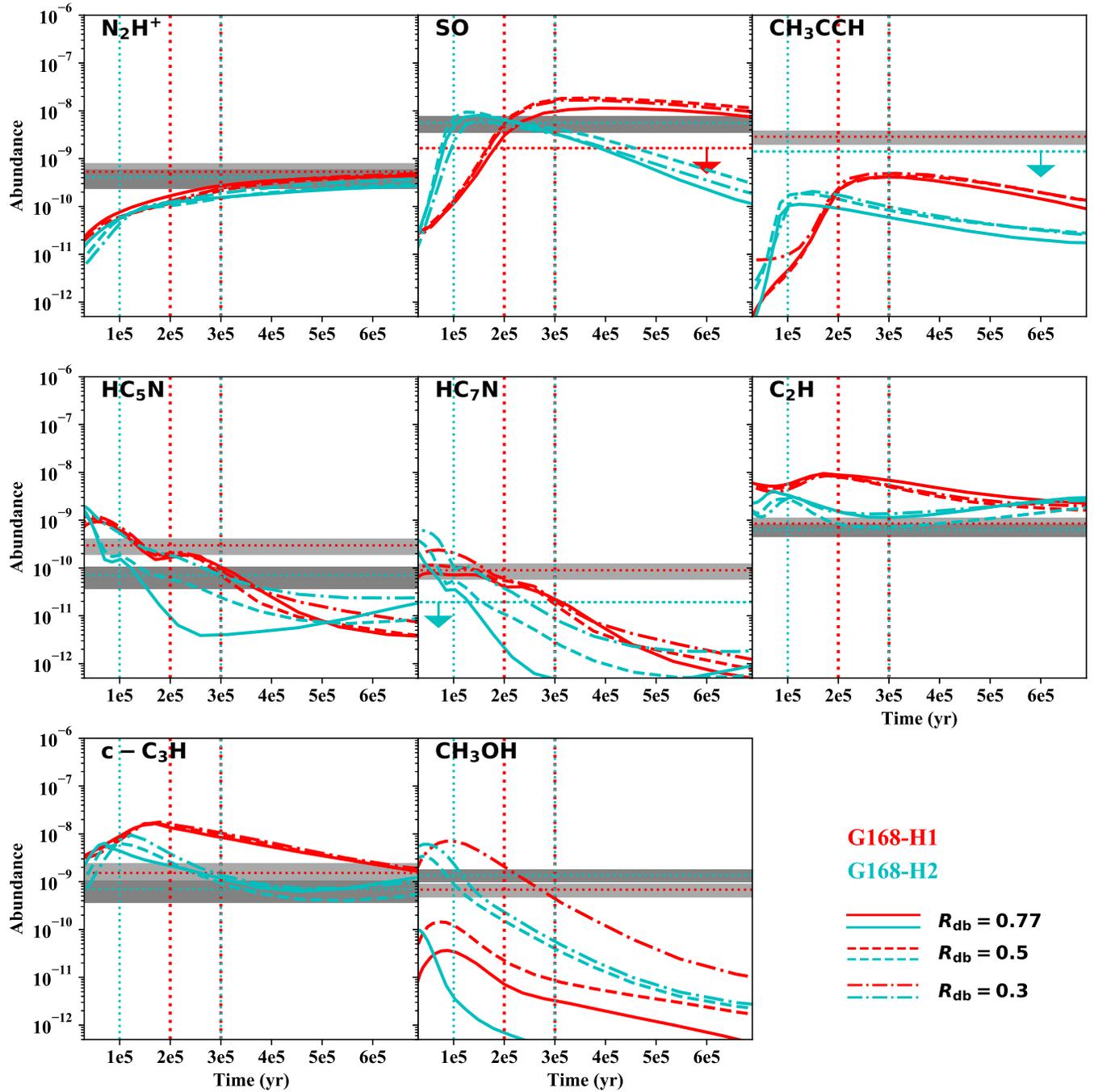}
\caption{Modeled abundance evolution tracks of molecules (marked by the black label in each panel) in G168-H1 (red) and G168H2 (cyan) with different diffusion-to-binding energy ratios ($R_{\rm db}$) of 0.3 (dashed-dotted line), 0.5 (dashed line), and 0.77 (solid line). The horizontal lines (dotted) with gray regions show the observed values and errors. The horizontal lines (dotted) with arrows show the upper limits derived from observations. The vertical dotted lines indicate the optimal chemical age ranges: (2$\sim$3)$\times10^{5}$ and (1$\sim$2)$\times10^{5}$\,yr for G168-H1 and G168-H2, respectively. The middle vertical lines (cyan and red) overlap with each other.}
\label{fig_1d}
\end{figure*}
Figure~\ref{fig_1d} shows the modeled abundance evolution tracks of molecules for G168-H1 and G168-H2, together with the observed abundances. From this figure, we see that all of the observed molecular abundances can be well reproduced by our models within one order of magnitude at the abovementioned chemical age ranges (vertical dotted lines). 

\ce{N2H+} is generally used as a chemical clock because its abundance increases with chemical age monotonously (see Figure~\ref{fig_1d}). However, neither the modeled nor observed abundances of \ce{N2H+} in G168-H1 and G168-H2 can be used to constrain their chemical ages owing to the small differences between the two cores.
However, the nondetection of SO provides a possible constraint on the chemical age of G168-H1. Compared to SO in G168-H2, which is roughly on the verge of depletion at the median chemical age of $\sim 1.5\times 10^5$\,yr, SO is less evolved in G168-H1, making it less likely to exhibit a detectable  abundance if G168-H1 has a chemical age of $\sim 2\times 10^5$\,yr. This trend of SO in the G168-H1 model is consistent with that found in other dark cloud models with densities of $\sim 2\times 10^4$ cm$^{-3}$ \citep[e.g.][]{Agundez13,Vidal17,Lass19}. This implies that the nondetection of SO in G168-H1 is due to the less-evolved state of G168-H1.
By changing the chemical age of G168-H1 to $\sim$2$\times$10$^{5}$ yr, their observed abundances can also be better reproduced by our models.
All these suggest that G168-H1 is less evolved compared with G168-H2 from the viewpoint of chemistry, which is consistent with their physical states indicated by their densities (see Table~\ref{tab_phy}).

According to our models with different diffusion-to-binding ratios ($R_{\rm db}$) of 0.3, 0.5, and 0.77 (see Figure~\ref{fig_1d}),
\ce{CH3OH} is more sensitive to the ratios. This demonstrates that it is mainly formed on dust grains
then released by cosmic-ray-induced desorption from small dust grains. The observations exhibit better agreement with models with fast surface reactions ($R_{\rm db}=0.3$ or 0.5), producing comparable abundances of \ce{CH3OH} within the reasonable chemical age ranges of G168-H1 and G168-H2.
Although \ce{HC5N} and \ce{HC7N} are also sensitive to the diffusion-to-binding ratios, they are mainly formed in gas-phase at low temperature via dissociative recombination reactions between molecular ions (\ce{C2H5N+} and \ce{C2H7N+}) and electrons \citep{Loison2014}.
The dependence of \ce{HC5N} and \ce{HC7N} chemistry on diffusion-to-binding energy ratios may mean that \ce{HC5N} and \ce{HC7N} formed in the gas phase are absorbed onto dust grains and sublimate into the gas phase by cosmic-ray-induced desorption mechanism.
For other carbon-chain molecules, our models also show higher abundances in G168-H1 than those in G168-H2, consistent with the observed trends. 

All the modeled features of the observed molecules support the discussion of their chemistry in section~\ref{chemical differences}.
The chemical difference between the two cores may be due to their different gas densities.
As we mentioned in section \ref{Introduction}, the G168-H2 core is more evolved and denser than G168-H1. Together with the results of observations and chemical models, we suggest that the chemical difference between the two cores may be due to their different gas densities and different evolutionary stages.

\section{Summary}
We carried out both observations and model calculations to investigate the chemical properties of the G168-H1 and G168-H2 cores in PGCC G168.
The main findings of this work are summarized as follows:
\begin{enumerate}
\item Through the single-position observations with the Tianma 65 m telescope, molecular lines of \ce{c-C3H}, \ce{H2CO}, and \ce{HC5N} were detected in both G168-H1 and G168-H2. Whereas \ce{HC7N} was detected only in G168-H1, and SO was detected only in G168-H2.
\item Mapping observations with the PMO 13.7 m telescope have detected CCH, \ce{N2H+}, and \ce{CH3OH} in both G168-H1 and G168-H2 but \ce{CH3CCH} only in G168-H1. CCH, \ce{N2H+}, and \ce{CH3CCH} have stronger emissions in the G168-H1 core than in the G168-H2 core, whereas \ce{CH3OH}, opposite the distribution of the other three species, presents stronger emission in G168-H2.
\item Carbon-chain species, such as CCH, \ce{CH3CCH}, \ce{HC5N}, and \ce{HC7N}, are less abundant in G168-H2 than in G168-H1. This is consistent with previous studies that carbon-chain molecules are deficient in more evolved prestellar cores.
\item When compared with other prestellar cores, lower \ce{N2H+} abundances in the prestellar G168-H2 core can be attributed to \ce{N2H+} depletion in the G168-H2 core. We suggest that the \ce{N2H+} depletion in G168-H2 is dominated by \ce{N2} depletion rather than destruction by CO.
\item By comparison with previous observations of carbon-chain molecules toward various objects, we suggest that the starless core G168-H1 may exhibit the properties of cold dark clouds, but the prestellar core G168-H2 may have an even lower abundance of carbon-chain molecules than general cold dark clouds do.
\item Chemical models have been conducted to explain the observed features in G168. Chemical modeling suggests that the chemical differences between G168-H1 and G168-H2 may be due to their different gas densities and different evolutionary stages.
\end{enumerate}

\section*{Acknowledgements}
This work has been supported by the National Key R\&D Program of China (No. 2017YFA0402701) and by the National Natural Science Foundation of China (grant Nos. 11373026 and 11433004), and the Joint Research Fund in Astronomy (U1631237) under cooperative agreement between the National Natural Science Foundation of China and Chinese Academy of Sciences, by the Top Talents Program of Yunnan Province (2015HA030). J.X. acknowledges support from FONDECYT grant 3170768 and the ``Light of West China'' program of Chinese Academy of Sciences (CAS).
Mengyao Tang is supported by China Scholarship  Council (CSC) and by Yunnan University’s Research Innovation Fund for Graduate Students.
We are thankful for the thoughtful suggestions from the anonymous referees that helped to improve our manuscript.

\appendix
\section{Physical Parameters}
\label{physical parameters}
Our gas-grain chemical model is a single-point model with fixed physical parameters.
The single-point models are set for the core centers of G168-H1 and G168-H2.
For the volume densities, we use the typical models of $n_{\rm H}=2\times 10^4$ and $n_{\rm H}=1\times 10^5$ cm$^{-3}$ for G168-H1 and G168-H2 respectively.
These values are in accordance with the reported results by \citet{Tang18}, which gives $ (1.9\sim4.6)\times10^4$ and $(0.49\sim 1.8)\times 10^5$ cm$^{-3}$ for G168-H1 and G168-H2, respectively, from calculations and fitting to column density maps derived from SPIRE continuum data by \citet{Marsh17}.
The gas temperature is fixed as 10 K according to the gas kinetic temperatures of 9.66 and 10.25 K for G168-H1 (R.A.=$4^{\rm h}18^{\rm m}33.1^{\rm s}$, Decl.=$28^{\circ}27^{\prime}11^{\prime\prime}$) and G168-H2 (R.A.=$4^{\rm h}18^{\rm m}39.1^{\rm s}$, Decl.=$28^{\circ}23^{\prime}29^{\prime\prime}$) respectively, deduced from NH$_{3}$ (1,1) and (2,2) by \citet{Seo15}.
We adopted the gas temperatures from \ce{NH3}(1,1) and (2,2) because they are excellent indicators of temperature \citep{Ho1983,Ungerechts1986,Danby1988}. We tested the effect of temperature on our model and found that the effects of temperature changes between 3 and 10 K were negligible.

Considering the dust component, the dust temperatures are set to 11.9 K and 13.2 K for G168-H1 and G168-H2, respectively \citep{Tang18}. We have tested models with well-coupled gas and dust temperatures in range of 10--20\,K which show that the effects of temperature on chemistry is negligible. Thus, the adopted dust temperatures from the literature is reasonable for our chemical models. Using the relation between extinction and \ce{H2} column density proposed by \citet{Guver09}, $A_{\rm V}=N(\ce{H2})/2.21\times 10^{21}$\,mag, the extinctions are estimated as 6.3 and 11.7\,mag for G168-H1 and G168-H2, respectively. Therefore, the visual extinction is fixed as a typical value of 10 mag for both cores. Meanwhile, we used the classical MRN dust size distribution \citep{Mathis77}, which can be expressed as
\begin{equation}
\frac{1}{n_{\rm H}}\frac{dn_d}{dr_d}= Cr_d^{-3.5}dr_d,\, 0.005 \leq r_d \leq 0.25\,\mu{\rm m},
\end{equation}
where $n_{\rm H}$ is the total number density of H atom, and $r_d$ is the dust radius. The grain constant C equals 10$^{-25.11}$cm$^{2.5}$, as is given by \citet{Weingartner01} for silicate grains.
\citet{Iqbal18} had shown that negligible chemical differences resulted from models with dust grains of different bin numbers of 10, 30, and 60.
Therefore, we use 10 bins in log-space to represent the dust size distribution to save computation cost.
Corresponding to the dust size distribution, size-dependent cosmic-ray-induced desorption is considered in the same way as  \citet{Iqbal18}.
Namely, the smallest dust grains heat to a peak temperature of $\sim$ 300 K \citep{Herbst06} via collision with a cosmic-ray particle, rather than the 70 K used in the model with single-size dust grains of $r_{d}$=0.1$\micron$ \citep{Hasegawa93}.
The radiation field is fixed as one in units of the FUV interstellar radiation field $\chi_0$ from \citet{Draine1978}.
The cosmic ionization rate is fixed as 3$\times$10$^{-17}$ s$^{-1}$.

\section{Best chemical age}
\label{chemical age}
\begin{figure}
\plotone{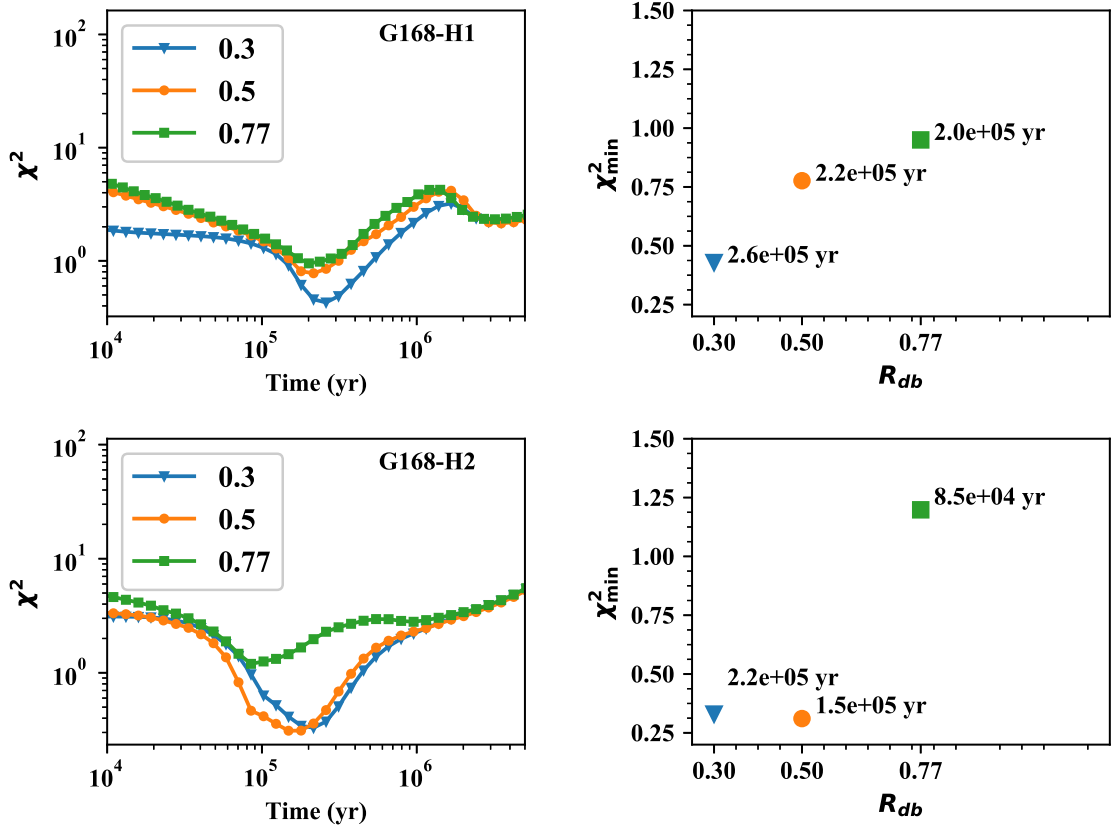}
\caption{Left panel: $\chi^2$ as function of time with varied values of $R_{\rm db}$ as 0.3 (triangle), 0.5 (circle) and 0.77 (square). Right panel: The minimum $\chi^2_{\rm min}$ as function of $R_{\rm db}$ showing with the same symbols as that in left panels. The labels show the corresponding best chemical ages. The upper and right panels are for G168-H1 and G168-H2, respectively.}
\label{fig_chi2}
\end{figure}
To determine the best chemical ages for G168-H1 and G168-H2, we use the chi-square method, which is expressed as
\begin{equation}
\chi^2(t) = \sum_{i=1}^{N} \left[ \log(X_{{\rm obs},i})-\log(X_{{\rm mod},i}(t))\right]^2,
\end{equation}
where $N$ = 8 is the number of observed species including the species only with upper limits determined. $X_{\rm obs}$ and $X_{\rm mod}$ are the observed and modeled abundances of species $i$, respectively.
Chi-square ($\chi^2$) as a function of chemical age with values of diffusion-to-binding energy ratio ($R_{\rm db}$) of 0.3, 0.5, and 0.77 is shown in the left panels of Figure~\ref{fig_chi2}.
The right panels show the minimum $\chi^2_{\rm min}$ as function of $R_{\rm db}$, with labels indicating the corresponding optimal chemical ages.
From this figure, we can see that models with different values of $R_{\rm db}$ have different optimal chemical ages. Because of the limited observed data, the dust properties cannot be constrained. Therefore, we use chemical age ranges to discuss the observed chemical features in this study. The chemical age ranges are (2$\sim$3)$\times 10^5$\,yr and (1$\sim$2)$\times 10^5$\,yr for G168-H1 and G168-H2, respectively.



\end{document}